\documentclass[twocolumn]{aastex63}
\usepackage{amsmath}
\usepackage{graphicx}
\usepackage{appendix}
\usepackage{comment}
\usepackage{tabularx}
\usepackage{booktabs}
\usepackage{comment}
\usepackage{xcolor}
\usepackage{hyperref}
\usepackage{lineno}
\usepackage{enumitem}

\newcommand{\h}{\mathbf{h}}
\newcommand{\Msun}{M\ensuremath{_\odot}}
\newcommand{\Rsun}{R\ensuremath{_\odot}}
\newcommand{\Zsun}{Z\ensuremath{_\odot}}
\newcommand{\Mzams}{\ensuremath{M_{\rm ZAMS}}}
\newcommand{\code}[1]{\texttt{#1}}
\newcommand{\mesa}{\code{MESA}}
\newcommand{\MESA}{\mesa}

\setlist[itemize]{noitemsep, topsep=0pt}
\newcommand{\tightitems}{
  \setlength{\topsep}{0pt}
  \setlength{\itemsep}{1pt}
  \setlength{\parsep}{0pt}
  \setlength{\parskip}{0pt}
}

\newlength{\apjcolwidth}
\newlength{\figwidth}
\newlength{\doublewide}
\begin{document}

\title{Resolving The Peak Of The Black Hole Mass Spectrum}

\shorttitle{Resolving The BH Mass Spectrum Peak}
\shortauthors{Farag et al.}

\author[0000-0002-5794-4286]{Ebraheem Farag}
\affiliation{School of Earth and Space Exploration, Arizona State University, Tempe, AZ 85287, USA}
\affiliation{Joint Institute for Nuclear Astrophysics - Center for the Evolution of the Elements, USA}

\author[0000-0002-6718-9472]{Mathieu Renzo}
\affiliation{Center for Computational Astrophysics, Flatiron Institute, New York, NY 10010, USA}

\author[0000-0003-3441-7624]{Robert Farmer}
\affiliation{Max-Planck-Institut f{\"u}r Astrophysik, Karl-Schwarzschild-Straße 1, 85741 Garching, Germany} 

\author[0000-0002-5107-8639]{Morgan T. Chidester}
\affiliation{School of Earth and Space Exploration, Arizona State University, Tempe, AZ 85287, USA}
\affiliation{Joint Institute for Nuclear Astrophysics - Center for the Evolution of the Elements, USA}

\author[0000-0002-0474-159X]{F.X.~Timmes}
\affiliation{School of Earth and Space Exploration, Arizona State University, Tempe, AZ 85287, USA}
\affiliation{Joint Institute for Nuclear Astrophysics - Center for the Evolution of the Elements, USA}

\correspondingauthor{Ebraheem farag}
\email{ekfarag@asu.edu}


\begin{abstract}

Gravitational wave (GW) detections of binary black hole (BH) mergers have begun to sample the
cosmic BH mass distribution. The evolution of single stellar cores predicts a gap in the BH mass distribution due
to pair-instability supernova (PISN). Determining the upper and lower edges of the BH mass gap can be useful for
interpreting GW detections from merging BHs. 
We use \MESA\ to evolve single, non-rotating, massive helium
cores with a metallicity of $Z = 10^{-5}$ until they either collapse
to form a BH or explode as a PISN without leaving a compact remnant. 
We calculate the boundaries of the lower BH mass gap for S-factors in the range S(300 keV) = (77,203) keV b, 
corresponding to the $\pm 3\sigma$ uncertainty in our
high resolution tabulated $^{12}$C($\alpha$,$\gamma$)$^{16}$O reaction
rate probability distribution function. We extensively test the temporal
and mass resolution to resolve the theoretical peak of the BH mass spectrum across the BH mass gap. 
We explore the convergence with respect to convective mixing and nuclear burning,
finding that significant time resolution is needed to achieve convergence.  
We also test adopting a minimum diffusion coefficient to help lower resolution models reach convergence. 
We establish a new lower edge of the upper mass gap as M\textsubscript{lower}
$\simeq$\,60$^{+32}_{-14}$\,\Msun\ from the $\pm 3\sigma$
uncertainty in the $^{12}\text{C}(\alpha, \gamma) ^{16}\text{O}$ rate. 
We explore the effect of a larger 3-$\alpha$ rate on the lower edge of the upper mass gap, 
finding M\textsubscript{lower} $\simeq$\,69$^{+34}_{-18}$\,\Msun. 
We compare our results with BHs reported in the Gravitational-Wave Transient Catalog.

\end{abstract}
\keywords{
Gravitational waves (678);
Black holes (162);
Nuclear astrophysics (1129);
Stellar physics (1621);
Core-collapse supernovae (304)
         }

\section{Introduction} \label{s.intro}

The BH initial mass function from single star evolutionary models predict three physics-driven transitions in the distribution.
In order of increasing mass, 
the first transition is set by the maximum possible neutron star mass,
the second by electron-positron pair production from energetic photons in the stellar interior of a massive star, and
the third by exothermic photodisintegration reactions which absorb enough energy in a high temperature stellar core 
for the model star to, once again, reach core collapse. 

\begin{figure*}[!tbh]
    \centering
    \includegraphics[width=0.9\textwidth]{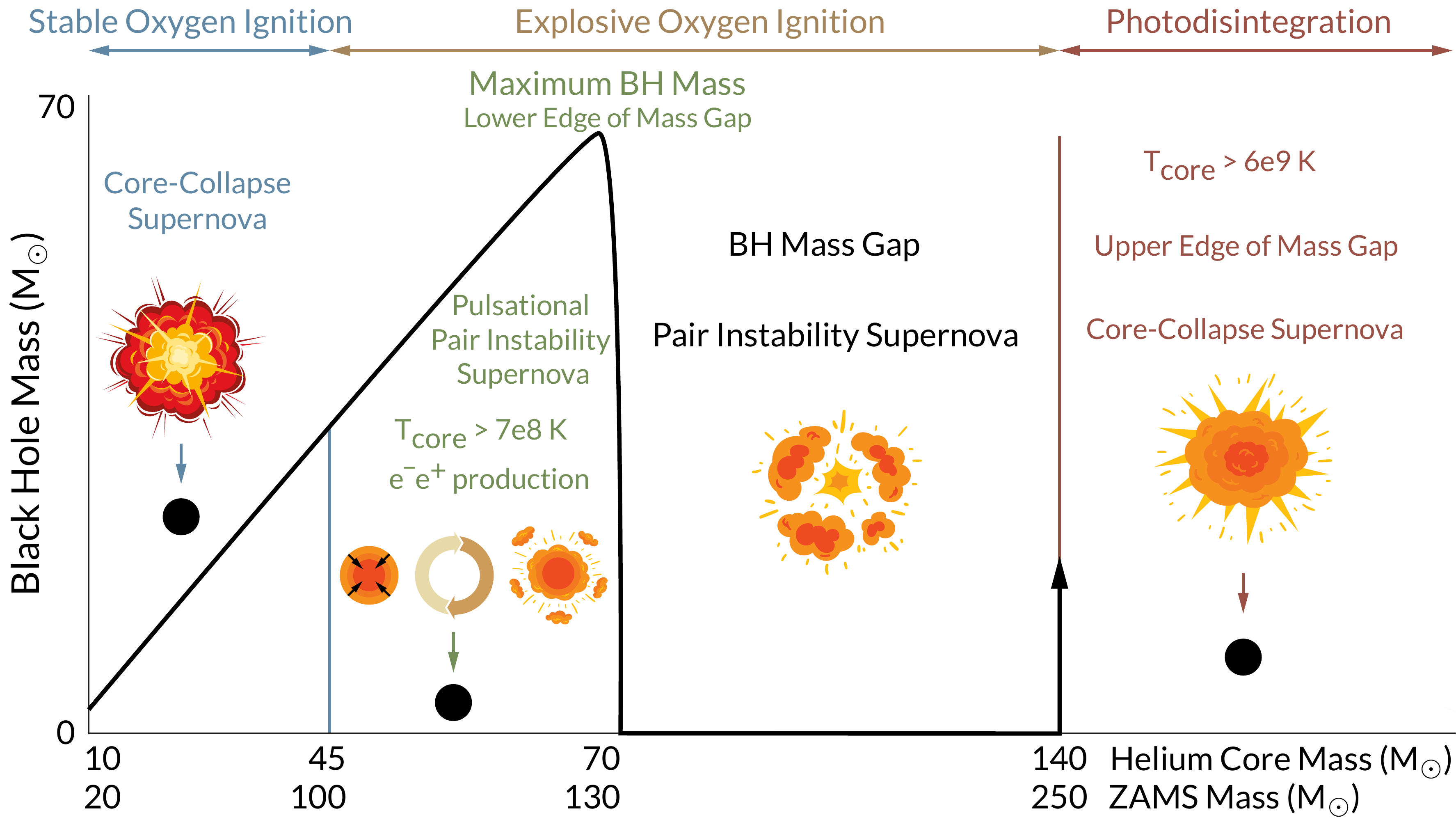}
    \caption{Illustration of a single star BH mass spectrum.}
    \label{fig:bh_mass_gap}
\end{figure*}

On the first transition,
the maximum observed masses of neutron stars include about
2.01\,\Msun\ for PSR J0348+0432 \citep{antoniadis_2013_aa},
about 2.08-2.14\,\Msun\ for PSR J0740+6620 \citep{cromartie_2020_aa,farr_2020_aa,riley_2021_aa,miller_2021_aa,fonseca_2021_aa}, and about 2.16\,\Msun\ for GW170817 \citep{abbott_2017_aj,abbott_2017_ak,rezzolla_2018_aa}.
Integration of the general relativistic
equations of hydrostatic equilibrium with a nuclear equation of state
predict a restricted range of allowed gravitational neutron star
masses, with currently favored equations of state giving
$\simeq$\,0.1--2.5\,\Msun\ \citep{banik_2014_aa,marques_2017_aa,ferreira_2021_aa,lattimer_2021_aa}.
The lower bound is the minimum stable neutron star mass \citep{oppenheimer_1938_aa,colpi_1989_aa,haensel_2002_aa,koliogiannis_2021_aa},
although a more relevant minimum mass stems from a neutron star's origin in a supernova
\citep{timmes_1996_ac,zhang_2008_ab,sukhbold_2016_aa,richers_2017_aa,sukhbold_2018_aa,ghosh_2022_aa,patton_2022_aa}
or accretion induced collapse of a white dwarf \citep{fryer_1999_aa, schwab_2021_aa, wang_2022_aa}.
Less massive neutron star masses can undergo explosive decompression \citep{page_1982_aa,colpi_1993_aa,sumiyoshi_1998_aa,nixon_2020_aa}
while more massive ones overcome the repulsive strong force and neutron degeneracy pressure to collapse into BHs.

In addition, there is an apparent paucity of observed BH candidates in the $\simeq$\,2.5--5\,\Msun\ mass range
\citep{bailyn_1998_aa,ozel_2010_aa,belczynski_2012_aa}, hinting at the possible existance of a 
contested ``lower mass gap'' \citep{farr_2011_aa,wyrzykowski_2020_aa,zevin_2020_aa,mandel_2022_ab}.
The ``lower edge'' of the lower mass gap is set by the maximum possible neutron star mass
while the ``upper edge'' is set by the minimum observed BH mass. 
The upper edge of the lower mass gap, if it exists, is currently uncertain as BHs found in this mass range may have been created by
changes in the growth time of convection during a SN explosion  \citep{fryer_2022_aa}, or by 
binary neutron star mergers \citep{thompson_2019_aa,abbott_2020_aa,gupta_2020_aa,yang_2020_aa,ye_2022_aa}.

The second and third transitions form the ``upper mass gap'', see Figure~\ref{fig:bh_mass_gap}.
Models with zero-age main sequence (ZAMS) masses \Mzams\,$\gtrsim$\,100\,\Msun\
reach central temperatures of $T_{\rm c}$\,$\gtrsim$\,7$\times$10$^{8}$ K. This allows
for the production of electron–positron pairs from photons, $\gamma$+$\gamma$\,$\rightarrow$\,$e^{-}+e^{+}$, 
that soften the equation of state \citep{fowler_1964_aa, barkat_1967_aa, rakavy_1967_ab}.
These models become dynamically unstable before core O-depletion as the pair production leads to
regions where the adiabatic index $\Gamma_1 \le$~4/3 \citep{fraley_1968_aa, ober_1983_aa, bond_1984_aa}.

The ensuing dynamical collapse results in explosive O-burning, with a variety of possible outcomes
\citep{glatzel_1985_aa, woosley_2002_aa, heger_2003_aa,takahashi_2018_aa, farmer_2019_aa,marchant_2019_aa, marchant_2020_aa,Renzo_2020b_aa,woosley_2021_aa, mehta_2022_aa,renzo_2022_aa}. The energy injected can cause a cyclic pattern of entering
the pair-instability region, contracting, burning, and expanding; eventually leading to
pulsational pair-instability supernovae (PPISN) and a BH remnant.
The energy injected by a single strong pulse can also unbind the model without leaving a BH remnant in a pair-instability supernova (PISN).
This transition to a PISN defines the ``lower edge'' of the upper mass gap.
Models with \Mzams\,$\gtrsim$\,250\,\Msun \ reach $T_{\rm c}$\,$\gtrsim$\,6$\times$10$^{9}$~K,
where endothermic photodisintegration reactions
absorb enough energy to prevent the star from unbinding \citep{bond_1984_aa,heger_2003_aa}.
This third transition defines the ``upper edge'' of the upper mass gap. 

The ZAMS mass of the second and third transitions depend on the He core-to-ZAMS mass mapping and thus the rotation, convective boundary mixing, and mass loss adopted \citep{vink_2015_cup,vink_2021_ras,higgins_2021_ras,woosley_2019_aa}. Since roughly half of all massive stars are found in binaries \citep{sana_2011_aa,sana_2012_aa}, a simple assumption is that binary interactions will strip the H envelope revealing bare He cores, before He burning commences. This can occur from Roche lobe overflow and or formation and ejection of a common envelope. All low-metallicity stars, including those with H-rich envelopes, are also expected to undergo PPISN/PISN and lose part or all of their envelope in the pulses \citep{woosley_2017_aa,2020_renzo_aa}. We are only interested in determining peak BH masses from GW sources in isolated systems, so we assume their progenitors are He cores. 

While the three transitions are set by nuclear and particle physics, the total number of BHs
as well as their distribution with mass is a consequence of stellar evolution.
A number of uncertain factors make a straightforward determination challenging.
For example, current estimates of the BH initial mass function from single stars
chiefly rely on parameterized explosion models \citep{spera_2015_aa,patton_2020_aa,zapartas_2021_aa,patton_2022_aa,fryer_2012_aa,fryer_2022_aa,renzo_2022_aa}. Models for the evolution from the ZAMS
also still rely on effective mixing length theories for convection, including ours.
Finally, BHs in either mass gap may occur through mechanisms other than those involving a single star.
Examples include
isolated binary star evolution \citep{de-mink_2016_aa,spera_2019_aa,breivik_2020_aa,Krzysztof_2020_aa,santoliquido_2021_aa,fuller_2022_aa,van-son_2020_aa},
dynamical formation in dense star clusters \citep{portegies-zwart_2000_aa,di-carlo_2020_aa,fragione_2022_aa,Renzo_2020b_aa},
mergers in higher multiplicity systems \citep{antonini_2012_aa,liu_2018_aa,hamers_2019_aa},
mergers of compact binaries in galactic nuclei \citep{oleary_2009_aa,bartos_2017_aa,wang_2021_aa},
and mass loss from putative progenitors above the PISN regime \citep[i.e., super-kilonova][]{siegel_2021_aa}.
Contributions from the different populations may be unraveled as the number of detected
gravitational wave events increase \citep{perna_2019_aa,zevin_2021_aa,renzo_2021_aa,mandel_2022_aa}.

This article focuses on the second transition, the lower edge of the upper mass gap.
The predicted BH mass at the lower edge from single naked He core models are generally robust
with respect to model uncertainties \citep{takahashi_2018_aa, farmer_2019_aa, marchant_2020_aa,farmer_2020_aa,2020_renzo_ras},
but depend sensitively on the $^{12}$C($\alpha$,$\gamma$)$^{16}$O reaction rate \citep{farmer_2020_aa}.
The masses and spins of merging binary BHs
from LIGO/Virgo/Karga \citep[LVK,][]{ligo-scientific-collaboration_2015_aa, acernese_2015_aa, akutsu_2021_aa} observations
probe the location of the lower edge of the upper mass gap.
The observed lower edge can then be used to place a constraint on the $^{12}$C($\alpha$,$\gamma$)$^{16}$O reaction rate
\citep{farmer_2020_aa}. However, we caution that LVK merging BHs might have a very complex astrophysical history
(e.g., binary evolution and stellar dynamics) which might blur the mass gap, and the mass gap may not appear at all with the LVK BHs.
Furthermore, we caution that there are competing effects at the lower edge of the mass gap (e.g., overshooting, collapse of the H-envelope)
that might have the same effect as varying $^{12}$C($\alpha$,$\gamma$)$^{16}$O reaction rates.

The main novelty of this article is a new effort, initiated by \citet{farmer_2020_aa} and refined in \citet{mehta_2022_aa},
aimed at showing convergence in the sequence of models used to define the lower edge of the upper mass gap
as a function of the He-burning reaction rates.
Section \ref{s.method} describes our models, Sections
\ref{s.bh_spectrum} and \ref{s.bh_massgap} describe our results, and
Section \ref{s.conc} summarizes our conclusions.

\section{Models}\label{s.method}

We use \MESA\ version r11701 to evolve single, non-rotating, massive
He cores with a metallicity of Z\,=\,10$^{-5}$ until they either
collapse to form a BH or explode as a PISN without leaving a compact
remnant. We adopt a low metallicity so that stellar winds are irrelevant \citep{farmer_2019_aa}, and to avoid the possible numerical
complications of resolving models with winds \citep{renzo_2017_aa}. Low metallicity environments are also likely to
form some of the most massive stellar mass BHs that can be detected as GW sources \citep{Michela_2021_aa,vink_2021_ras,mandel_2022_ab,spera_2022_aa}. 
Metallacities as low as Z\,=\,0.02\,\Zsun\ are enough to yield
final He core masses of up to 140\,\Msun\ when adopting recently
updated and physically motivated Wolf-rayet mass loss schemes
\citep{higgins_2021_ras}.

We use the same \MESA\ inlists and run\_star\_extras.f90 used in
\citet{farmer_2020_aa} and \citet{mehta_2022_aa} to calculate BH mass
spectrum across the lower edge of the upper mass gap. We also use a
21-isotope nuclear network with nuclear reactions rates from the
NACRE compilation \citep{angulo_1999_aa} and JINA reaclib database \citep{cyburt_2010_ab}. The tabulated
$^{12}$C($\alpha$,$\gamma$)$^{16}$O
reaction rates were originally provided by \citet{deboer_2017_aa}
and refined in \citet{mehta_2022_aa}.
A total of 1750 models were run that consumed $\simeq$\,3,000,000 core-hours.
Full details of the \MESA\ models and reaction rate files to reproduce our results are
available at doi:\dataset[https://doi.org/10.5281/zenodo.6930577]{[https://doi.org/10.5281/zenodo.6930577}.

Preceding any individual pulse, we use \MESA's implicit hydrodynamic solver.
As a model evolves into the pair-instability region we switch to \MESA's Hartan-Lax-van Leer-Contact hydrodynamic solver \citep[HLLC,][]{1994_Toro_ams, paxton_2018_aa} to resolve shocks during
the dynamic phase of evolution. The switch generally occurs when  central temperatures exceed $10^{9}$ K and the volumetric
pressure-weighted average adiabatic index $\langle\Gamma_{1}\rangle-4/3 < 0.01$
\citep{stothers_1999_aa,marchant_2019_aa,farmer_2019_aa}. We then follow the core as it gravitationally unbinds on the first pulse as a PISN,
or contracts and rebounds repeatedly until all shocks have reached the surface of the model
\citep{yoshida_2016_aa,woosley_2017_aa,marchant_2019_aa,2020_renzo_ras}.
Individual pulses can eject between $\simeq$\,0.01\,\Msun\ and $\simeq$\,30\,\Msun\ with surface velocities up to\,20000 km s$^{-1}$. 

Once hydrostatic equilibrium is reached again, the unbound material is removed,
the remaining mass is relaxed to a new stellar model with an identical entropy
and chemical profile. This is possible since the removed mass is already moving at speeds beyond both the escape velocity and the surface sound speed, so no back-reaction on the remaining material is expected \citep{marchant_2019_aa}.
The evolution then continues with the implicit solver.
The new model contracts as it loses energy due to radiation and neutrino emission
until it undergoes an additional pulse or collapses to form a BH.
At the onset of core collapse (CC), the HLLC solver is activated to capture the dynamics of the infalling core.
We define CC to occur when any part of the model begins collapsing with velocities $v>$\,8000 km s$^{-1}$,
so as to capture any pulse that might be ejected during CC. The mass of the BH
is defined as the mass of bound material at CC \citep{2020_renzo_ras}.
These BHs masses are upper limits due to uncertainties in BH formation
\citep{fryer_2001_aa, branch_2017_aa, uchida_2019_aa, mandel_2022_aa, renzo_2022_aa}
and weak shock generation \citep{nadezhin_1980_aa,fernandez_2018_aa,ivanov_2021_aa,rahman_2022_aa}.

\section{Convergence of the Peak BH Mass Spectrum}\label{s.bh_spectrum}
Mapping the final evolutionary phases of massive stars remains challenging given the difficulty in 
resolving the interplay between convection, nuclear burning, rotation, convective-core overshooting, radiative transport, internal waves, 
mass loss eruptions, and binary interactions \citep{quataert_2012_aa,shiode_2014_aa,matzner_2021_aa,vink_2021_ras,Tanikawa_2021_aa,jacobson-galan_2022_aa,wu_2022_aa}.
All else equal, mass resolution is an important consideration to accurately control for changes in stellar structure \citep{farmer_2016_aa}. Assessing the sensitivity to mass resolution is a necessary practice to ensure robust predictions. The evolution of PPISN stars are characterized by short episodes of strong nuclear burning in their core rapidly changing their composition and density. We will control changes in the central density for temporal resolution. PPISN stars also eject large amounts of material at and beneath escape velocity. We will control the spatial mesh resolution of our models to resolve this material.

The goal of this section is to assess the numerical convergence the \MESA\ models used to determine the lower edge of the upper mass gap
with respect to spatial and temporal resolution.
During hydrostatic phases of evolution when the implicit solver is active,
we control the number of cells with \code{max\_dq}, the maximum fractional mass in a cell.
That is, the minimum number of cells is 1/\code{max\_dq}.
During the dynamic phases of evolution when the HLLC solver is active,
we control the number of cells with \code{split\_merge\_amr\_nz\_baseline}.
For both phases of evolution we primarily control the timestep with \code{delta\_lgRho\_cntr\_limit} which limits the timestep such that the change in the logarithm of the central density
is less than a specified fraction.

\begin{figure}[!htb]
    \centering
    \includegraphics[width=3.2in]{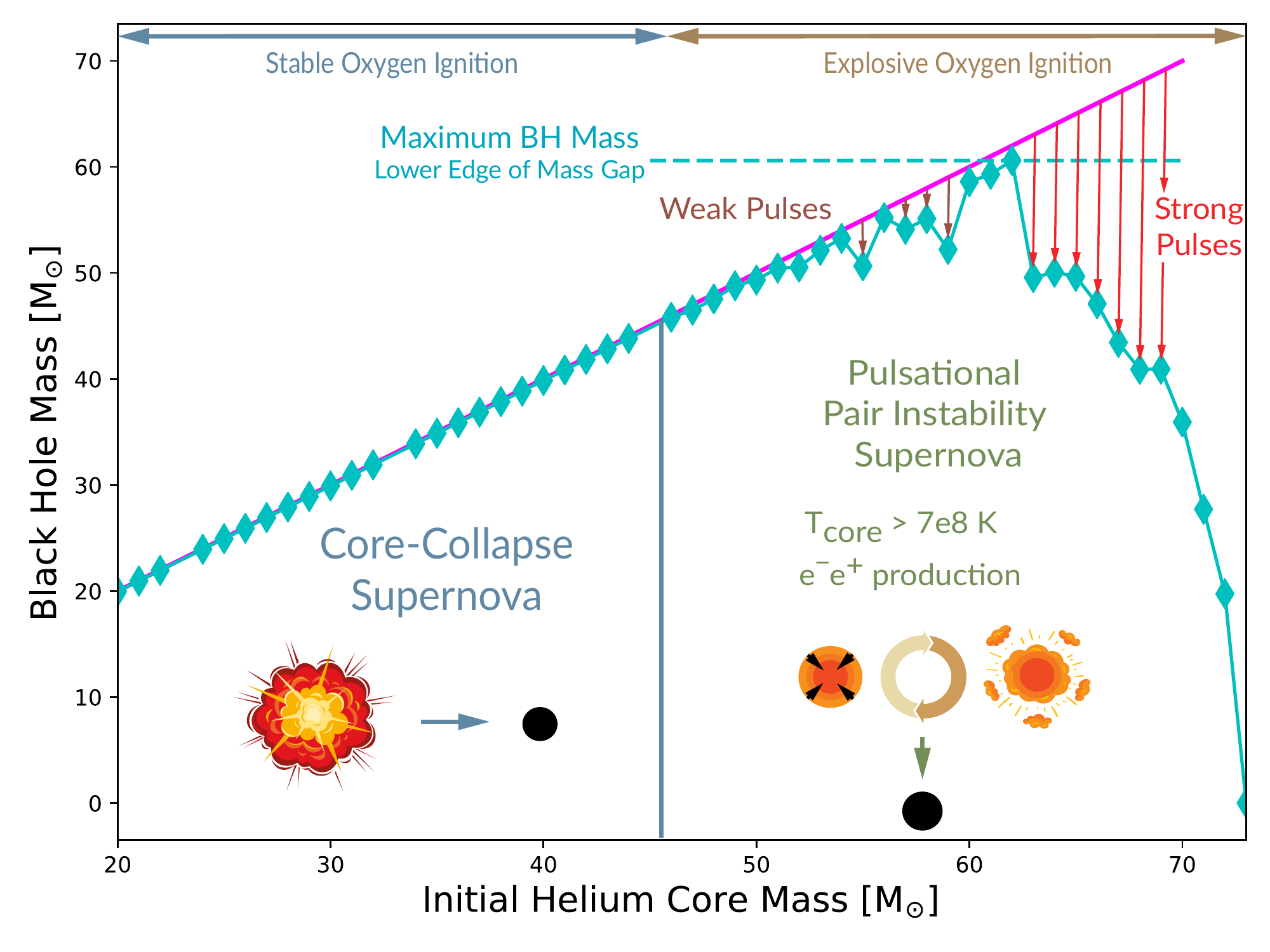}
    \caption{As in Figure~\ref{fig:bh_mass_gap}, but with an example of a calculated BH mass spectrum (teal line and diamonds). A magenta line marks $M_{\rm BH}$ \,=\, $M_{\rm He}$.
The peak BH mass is labeled. Smaller masses, just left of the peak, show weak pulses. Larger masses, just right, show strong pulses.}
    \label{fig:bh_mass_spec}
\end{figure}

In the following sections we indicate a ``strong'' pulse as a single pulse that arises from $\langle\Gamma_{1} - 4/3\rangle
<0$ and results in the ejection of $>$0.1\,\Msun\ of material. We indicate ``weak'' pulses as any rebounding pulse formed while the model relaxes from a strong pulse. Weak pulses also arise at or just above $\langle\Gamma_{1} - 4/3\rangle =0$
from dynamical behavior that is not strong enough to remove $>$0.1\,\Msun\ of material with a single shock, yet can remove
$\gg$\,0.1\,\Msun\ in a series of several or significantly more weak pulses which compound shocks near the surface. 
Strong pulses typically develop on a timescale of $\leq 50$ seconds, while typical periods for
the contraction and bounce of a single weak pulse can range between, but are not limited to, 10$^{2}$--10$^{3}$ seconds. 
In this paper, a PPISN is any model in which a single strong pulse or a series of weak pulses is 
able to remove $\geq$\,0.1\,\Msun\ of material from the surface. Figure \ref{fig:bh_mass_spec} illustrates a sample BH mass spectrum with weak and strong pulses.

\subsection{Global mixing floors}\label{s.mixing_floors}

\begin{figure*}
    \centering
    \includegraphics[width=7.2in]{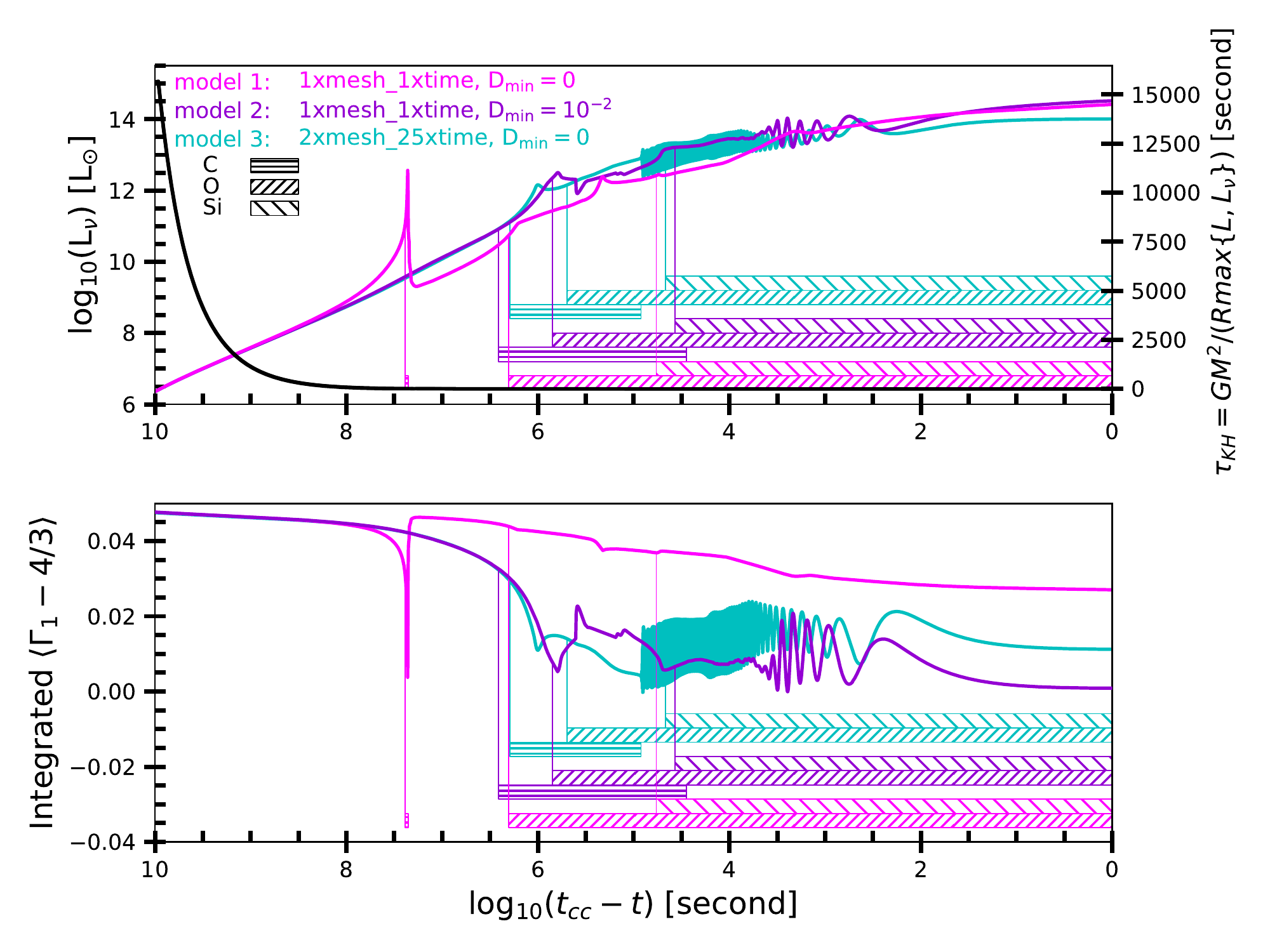}
    \caption{Both panels share the same x-axis, the logarithmic time to core-collapse (CC) in seconds. The top panel shows the neutrino luminosity and Kelvin-Helmholtz timescale, normalized to account for the energy lost to neutrinos, which dominates over photon cooling in the evolutionary phases shown. The integrated $\langle\Gamma_{1}-4/3\rangle$ is shown in the bottom panel to illustrate when the models are evolving dynamically.  Hatched regions indicate where C-shell, O-core, and Si-core burning take place.}
    \label{fig:fig1}
\end{figure*}

\begin{figure*}
    \centering
    \includegraphics[width=7.2in]{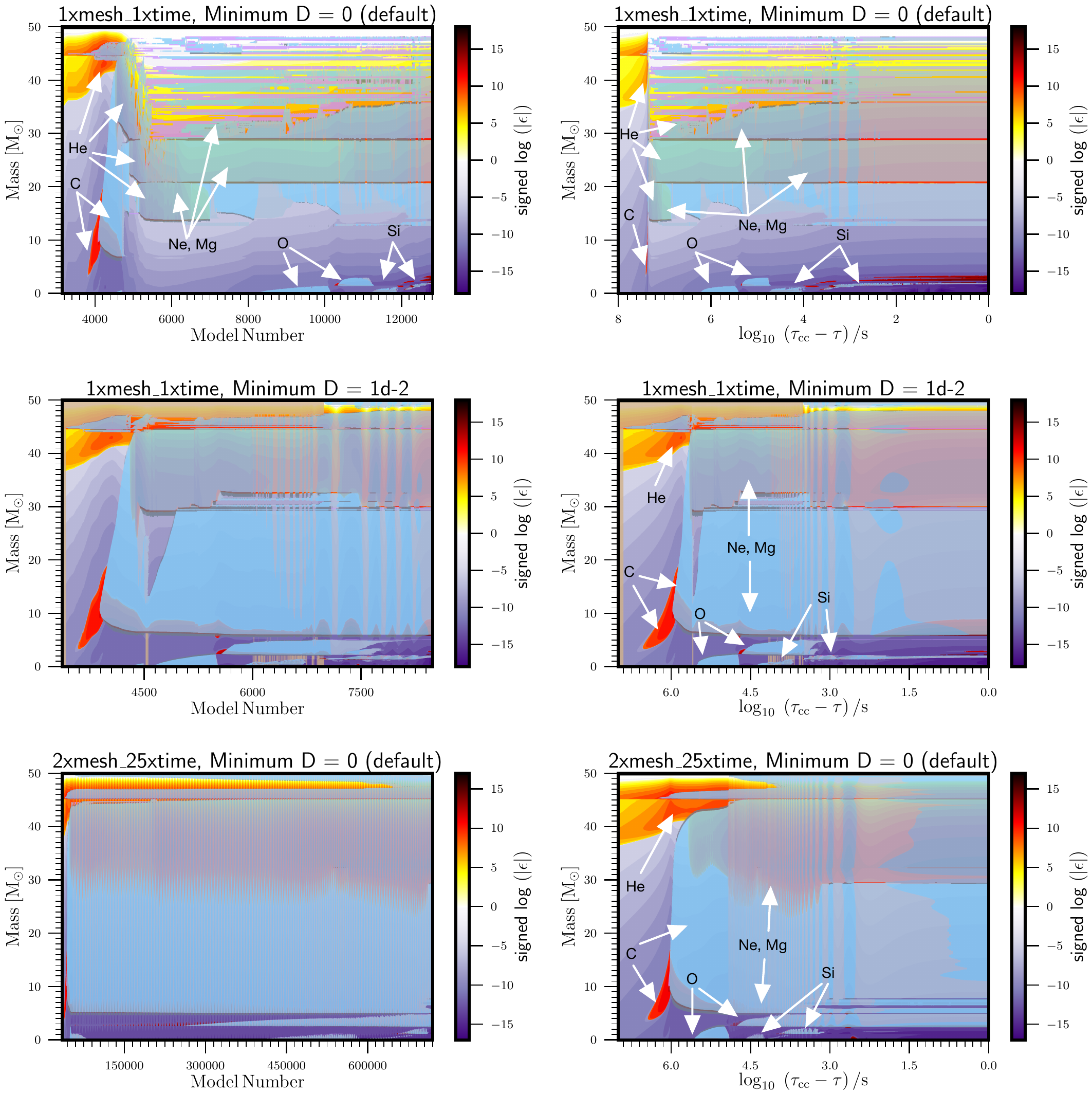}
    \caption{Difference in convection and nuclear burning between three stellar models in Section \ref{s.mixing_floors}. Figures in the left column are shown versus model number from He-burning to CC.  Figures in the right column are shown versus logarithm of the time to CC, beginning at the onset of shell C-burning. The different fuels burning are labelled. Each figure shows the signed logarithm of the net specific power, $\rm{sign}\left(\epsilon_{\rm{nuc}}-\epsilon_{\nu}\right)\log_{10}\left(\rm{max}\left(1.0,|\epsilon_{\rm{nuc}}-\epsilon_{\nu}|\right)\right)$, where $\epsilon_{nuc}$ is the specific energy generation rate and $\epsilon_{\nu}$ is the specific energy loss from neutrinos. Purple regions denote strong neutrino cooling and red regions denote regions of strong nuclear burning. Light blue regions indicate standard mixing length convection,  grey regions indicate convection with reduced mixing because of shorter convective timescales, and for model 2 beige-orange regions (e.g. near the surface) indicate where the diffusive mixing coefficient is set to D $=10^{-2}$ cm\textsuperscript{2}/s.}

    \label{fig:fig2}
\end{figure*}

Because of the finite-size of the Lagrangian mesh representing
a star in numerical models, sharp gradients (e.g., in chemical composition
or thermal properties) can be hard to resolve. This can lead to
  unphysical ``spikes'' that can have an oversized and
  resolution-dependent effect on stellar processes depending on those
  gradients (e.g., mixing instabilities). To mitigate this effect, 
a global minimum chemical mixing diffusion coefficient $D_{\rm min}$ can be used. 
Physically, $D_{\rm min}$ may be interpreted as a small amount of mixing arising from small scale perturbations not captured in one-dimensional models. The global
evolution of a model is unaffected when the associated diffusive
mixing timescale $\tau = L^{2}/D_{\rm min}$, where $L$ is the characteristic length-scale over which the mixing is active, is significantly
longer than the lifetime of the stellar model. Choosing
$D_{\rm min}$\,=\,10$^{-2}$ cm$^2$ s$^{-1}$, \code{min\_D\_mix} in
\MESA, and a typical PPISN radius of $R$\textsubscript{He}\,=\,1-10\,\Rsun, the global
mixing timescale is $\tau \simeq 10^{17}$ yr. A 50\,\Msun\ He core has
a lifetime of $\simeq$\,10$^{5}$ yr, about 12 orders of magnitude
smaller than this global mixing timescale. Thus, we do not anticipate a significant \emph{global} impact with this value of $D_{\rm min}$, while it smooths \emph{local} composition gradients. Nevertheless, models at the boundaries of different physical behaviors (e.g., core-collapse or pulsational pair-instability) may differ qualitatively.

Figure \ref{fig:fig1} and \ref{fig:fig2} illustrate the difference in evolutionary
behavior between three $M_{\rm He}$\,=\,50\,\Msun\ models from the advent of shell C-burning to CC. The thermal evolution of these models is dominated by neutrino cooling and therefore all three models remain in local
thermodynamic equilibrium. Figure \ref{fig:fig1} compares the luminosities, timescales, energies, and integrated $\langle\Gamma_{1}-4/3\rangle$
of three 50\,\Msun\ He core models as they evolve to CC. Figure \ref{fig:fig2} displays the convective and nuclear burning behavior of these models.
Model 1 uses the same mass and temporal resolution as \citet{farmer_2020_aa}:
\begin{itemize}\tightitems
\item[] \code{max\char`_dq}\,=\,1d-3
\item[] \code{split\_merge\_amr\_nz\_baseline}\,=\,6000
\item[] \code{delta\_lgRho\_cntr\_limit}\,=\,2.5d-3
\item[] \code{min\_D\_mix}\,=\,0.
\end{itemize}
Model 1 uses
$\simeq$\,2000--3000 cells during the hydrostatic phases,
$\simeq$\,2500--3500 cells during the dynamic phases,
and $\simeq$\,13000 timesteps to reach CC.
Model 2 is the same as Model 1, but sets \code{min\_D\_mix}=\,1d-2.
Model 2 uses
$\simeq$\,2000--3000 cells during the hydrostatic phases,
$\simeq$\,2500--4200 cells during the dynamic phases,
and $\simeq$\,9000 timesteps to reach CC.
Model 3 is the same as Model 1, but with about twice the mass resolution and 25 times the temporal resolution:
\begin{itemize}\tightitems
\item[] \code{max\_dq}\,=\,2d-4
\item[] \code{split\_merge\_amr\_nz\_baseline}\,=\,12000
\item[] \code{delta\_lgRho\_cntr\_limit}\,=\,1d-4
\item[] \code{min\_D\_mix}\,=\,0.
\end{itemize}
Model 3 uses
$\simeq$\,3500--4000 cells during the hydrostatic phases,
$\simeq$\,5000--10,000 cells during the dynamic phases,
and $\simeq$\,740,000 timesteps to reach CC.

In model 1, 2, and 3 off-center C-ignition occurs at log$(t_{cc}-t)\simeq 7.4$, log$(t_{cc}-t)\simeq
6.5$, log$(t_{cc}-t)\simeq 6.4$. In model 1, the chemical evolution of the core is interrupted by a dynamical episode at
log$(t_{cc}-t)\simeq 7.3$. Shell C-burning is an insufficient energy source to combat the dynamical instability as only a fraction of the $^{12}$C in the shell in burned before a $\Gamma_{1}$ instability leads to a core contraction. During this contraction the He-burning shell propagates inward, mixing and burning alongside Ne, Mg, and Si in the O-rich shell, stabilizing the core until CC.

Models 2 and 3 smoothly evolve through shell C-burning and undergoing weak dynamical behavior, and a
steady decrease in total energy as their cores smoothly evolve toward CC. With $D_{\rm min}$\,=\,10$^{-2}$ cm\textsuperscript{2} s$^{-1}$, model 2 experiences minimal mixing between the He-rich shell and the C-O core commensurate with model 3. Models 2 and 3 both smoothly burn through most of the $^{12}$C in the C-O shell while their cores burn through oxygen. This convective C-burning shell stabilizes the star long enough to prevent a dynamical contraction that disrupts core burning, unlike model 1. When the mass fraction of carbon, $X$($^{12}$C), in the shell drops below $\sim$\,10$^{-2}$, shell burning is no longer strong enough to prevent $\Gamma_{1}$ instability from pair-production. The shell begins to contract outward-in rebounding off the core from brief episodes of explosive O-burning. These explosive burning episodes drive weak shocks outward, momentarily stabilizing the shell until the shocks reach the surface and the shell contracts again. During each contraction, the temperatures and densities increase causing Ne and Mg burning in the outer shell to reach deeper within the model with every contraction, and recede back toward the surface during each pulse. As the core evolves further toward CC, Si-burning contributes to each episode. These pulsations can be seen in the oscillations of $\langle\Gamma_{1}-4/3\rangle$ and the oscillatory convection in Figures \ref{fig:fig1} and \ref{fig:fig2}. Model 2 manages to resolve a few pulsations before core-collapse, each ejecting a small amount of mass, resulting in M\textsubscript{eject}$\simeq$\,0.01\,\Msun. The higher resolution of model 3 is able to resolve $>140$ pulses before reaching CC, each pulse releasing a small amount of mass, resulting in  M\textsubscript{eject}$\simeq$\,0.45\,\Msun.

Adopting $D_{\rm min}$\,=\,10$^{-2}$ cm\textsuperscript{2} s$^{-1}$ improves the rate of numerical convergence of low resolution models akin to a slight increase in temporal resolution. While model 2 can resolve core burning reasonably well, the timesteps are too large to fully resolve nuclear burning in the shell or the coupling between the core and shell that generates oscillatory burning in the core during later stages, as in model 3.  Because our timesteps are primarily limited by \code{delta\_lgRho\_cntr\_limit}, small changes in central density resulting from the dynamical contraction between each core and shell burning episode can be fully resolved in model 3 but not in models 1 or 2. Only the strongest, largest core-shell contractions can be resolved at the lower resolution of model 2. Since each contraction does not change the core density at an appreciably large level, timesteps taken by model 2 skip or smooth over many of these pulsations. This suggests that $D_{\rm min}$\,=\,10$^{-2}$ cm\textsuperscript{2} s$^{-1}$ improves the rate of convergence of models run with lower resolutions, but cannot fully recover the convective behavior generated by models which use significantly greater temporal resolutions. Next we continue our exploration of $D_{\rm min}$\,=\,10$^{-2}$ cm\textsuperscript{2} s$^{-1}$  at higher resolutions to better characterize the impact across the entire BH mass spectrum.

\begin{deluxetable*}{lllllclllc}[!htb]
  \tablenum{1}
  \tablecolumns{10}
  \tablewidth{\doublewide}
  \tablecaption{
   Initial He core mass, composition at He depletion, and resolution controls for the models producing the peak BH mass below the upper mass gap.\label{tab:table1}
  }
  \tablehead{
    \colhead{\code{Name}} &
    \colhead{\code{min D}} &
    \colhead{\code{max\_dq}} &
    \colhead{\code{split\_nz}} & \colhead{\code{$\delta$\textsubscript{log$\rho_{c}$}}} &
    \colhead{\code{M\textsubscript{He}}(\Msun)} &
    \colhead{\code{M\textsubscript{CO}\textsubscript{\_He}}(\Msun)} &
    \colhead{\code{X(\textsuperscript{12}C)\textsubscript{He}}} &
    \colhead{\code{Pulses}}  &
    \colhead{\code{M\textsubscript{BH}}(\Msun)}
    }
\startdata
        \code{1m\_1h\_1t$^{a}$ \citep{farmer_2020_aa}} & 0 & 1d-3 & 6000 & 2.5d-3 & 53 & 47.59 & 0.1807 & 0 & 52.77 \\
        \code{2m\_1h\_2p5t \citep{mehta_2022_aa}}      & 0 & 5d-4 & 6000 & 1d-3   & 60 & 54.51 & 0.1699 & 1 strong, 6 weak & 59.17 \\
        \code{1m\_1h\_2p5t}   & 0    & 1d-3 & 6000  & 2.5d-3 & 58 & 52.53 & 0.1730 & 1 strong, $\sim6$ weak & 57.40  \\
        \code{2m\_2h\_2p5t}   & 0    & 5d-4 & 12000 & 2.5d-3 & 54 & 53.76 & 0.1757 & 4 strong, $>50$ weak & 52.3 \\
        \code{2m\_2h\_5t}     & 0    & 5d-4 & 12000 & 5d-4   & 60 & 54.50 & 0.1696 & 1 strong, $\sim7$ weak  & 59.07  \\
        \code{2m\_2h\_5t\_D}  & 1d-2 & 5d-4 & 12000 & 5d-4   & 60 & 54.15 & 0.1694 & 1 strong, $\sim7$ weak & 59.23  \\
        \code{2m\_2h\_25t}    & 0    & 5d-4 & 12000 & 1d-4   & 62 & 56.24 & 0.1670 & 1 strong, $\sim9$ weak & 60.61  \\
        \code{5m\_2h\_2p5t}   & 0    & 2d-4 & 30000 & 1d-3   & 55 & 48.92 & 0.1687 & 0 & 54.76  \\
        \code{5m\_5h\_2p5t}   & 0    & 2d-4 & 30000 & 1d-3   & 55 & 48.92 & 0.1687 & 0 & 54.76  \\
        \code{5m\_5h\_5t}     & 0    & 2d-4 & 30000 & 5d-4   & 57 & 50.34 & 0.1688 & 1 strong, $\sim10$ weak & 53.43  \\
        \code{5m\_5h\_5t\_D}  & 1d-2 & 2d-4 & 30000 & 5d-4   & 57 & 50.28 & 0.1719 & 1 strong, $\sim5$ weak & 55.41  \\
        \code{5m\_2h\_10t}    &  0   & 2d-4 & 12000 & 2.5d-4 & 62 & 56.29 & 0.1672 & 1 strong, $\sim10$ weak & 60.17   \\
        \code{5m\_2h\_10t\_D} & 1d-2 & 2d-4 & 12000 & 2.5d-4 & 62 & 56.39 & 0.1673 & 1 strong, $\sim10$ weak & 60.07  \\
        \code{5m\_2h\_25t}    & 0    & 2d-4 & 12000 & 1d-4   & 62 & 56.17 & 0.1674 & 1 strong, $\sim10$ weak & 60.55  \\
        \code{5m\_5h\_25t}    & 0    & 2d-4 & 30000 & 1d-4   & 62 & 56.17 & 0.1674 & 1 strong, $\sim10$ weak & 60.32  \\
        \code{5m\_5h\_25t\_D} & 1d-2 & 2d-4 & 30000 & 1d-4   & 62 & 56.17 & 0.1674 & 1 strong, $\sim10$ weak & 60.37  \\
  \enddata

 \tablenotetext{a}{In \citet{farmer_2020_aa} the default values are \code{max\_dq}\,=\,1d-3,
\code{split\_merge\_amr\_nz\_baseline}\,=\,6000, and
\code{delta\_lgRho\_cntr\_limit}\,=\,2.5d-3. All models in this table
are scaled multiples of these values. For example, \code{2m\_2h\_2p5t}
represents 2x the number of cells during the hydro-static phase, 
2x the number of cells during the hydrostatic and hydrodynamic phase, 
and 2.5x times the temporal resolution during all phases of evolution. 
\code{split\_merge\_amr\_nz\_baseline}
is labeled as \code{split\_nz} and \code{delta\_lgRho\_cntr\_limit} is labeled as
$\delta$\textsubscript{log$\rho_{c}$}.
M\textsubscript{CO\_He} and
X(\textsuperscript{12}C)\textsubscript{He} are respectively the C-O
core mass and central $^{12}$C mass fraction at core He-depletion. 
}
\end{deluxetable*}

\subsection{Resolution Testing The Single Star BH Mass Spectrum}\label{s.bh_res_spectrum}
\begin{figure*}
    \centering
    \includegraphics[width=7in]{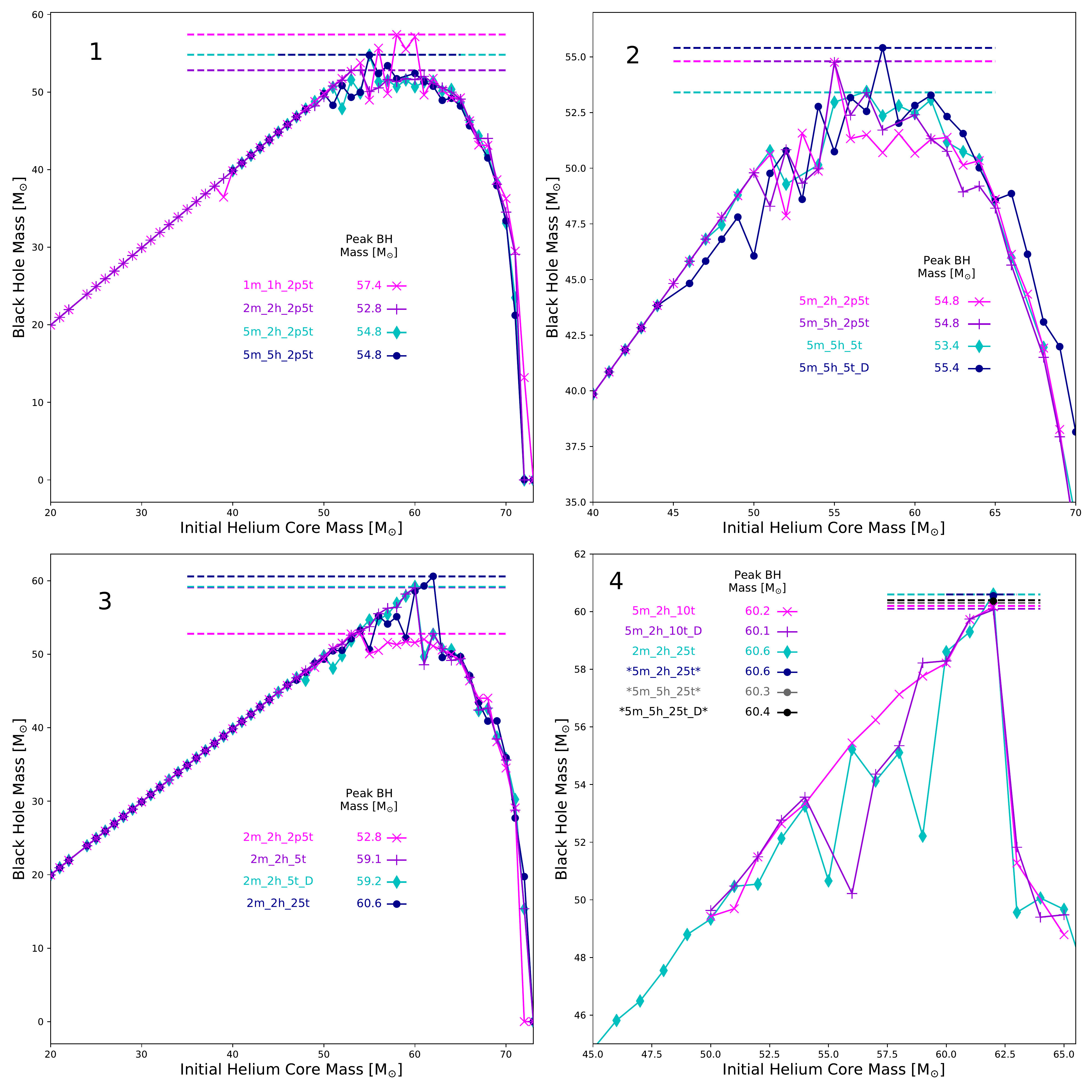}
    \caption{BH mass spectrums for the resolutions in Table \ref{tab:table1}. All panels share a similar x-axis, Initial He Core Mass M\textsubscript{He} in \Msun, and y axis, BH Mass in \Msun. Resolutions only shown at their respective peak BH mass are labeled with asterisks. Horizontal dashed and solid lines indicate the location of the peak BH mass for each resolution.}
    \label{fig:fig3}
\end{figure*}

In \citet{farmer_2019_aa}, \citet{marchant_2019_aa}, \citet{Renzo_2020b_aa} and \citet{mehta_2022_aa}, brief explorations of enhanced mass and temporal resolution were conducted, but further work was needed to assess the convergence of these models across the entire BH mass spectrum. In this section we take on the challenge of exploring the numerical convergence of \MESA\ stellar models computed across the BH mass spectrum. The goal of this section is to achieve convergence in the peak BH mass at the $<$\,1\,\Msun\ level. We focus on resolution testing the BH mass spectrum for He cores with the $\sigma$\,=0 $^{12}$C($\alpha$,$\gamma$)$^{16}$O reaction rate provided by \citet{deboer_2017_aa} and updated to 2015 temperature points in \citet{mehta_2022_aa}, corresponding to an approximate astrophysical S-factor S(300 keV) = 140 keV b. We also investigate the inclusion of a minimum diffusive mixing floor $D_{\rm min}$ on the BH mass spectrum.

Our testing consists of 16 different model resolutions, each modified using some variation of \code{max\_dq}, \code{split\_merge\_amr\_nz\_baseline}, and \code{delta\_lgRho\_cntr\_limit} to control for mass and temporal resolution. In Table \ref{tab:table1} we summarize the adopted resolution and associated peak BH spectrum mass. In Figure \ref{fig:fig3} we plot the BH mass spectrum for each resolution listed in Table \ref{tab:table1}. The baseline resolution of \code{1m\_1h\_1t} represents the resolution adopted in \citet{farmer_2020_aa}. All other models are scaled up from \code{1m\_1h\_1t}. Generally models with \code{1m}, \code{2m}, and \code{5m} have between 2000-3000, 3500-4000, and 7000-8000 cells respectively during the hydrostatic phase of evolution. Models with \code{1h}, \code{2h}, and \code{5h} have between 2500-4500, 5000-10000, and 12000-25000 cells respectively during the hydrodynamic phase of evolution. The number of cells typically increases from the lower bound to the higher bound as a model evolves and the average density of the model generally increases (except during pulses). Models with timesteps set by \code{1t}, \code{2p5t}, \code{5t}, \code{10t}, and \code{25t} generally take between 8000-40000,  15000-150,000, 20,000-200,000, 40,000-500,000, and 100,000-1,800,000 timesteps  to reach CC depending on the number of pulses encountered during the evolution. Models that produce the peak of the BH mass spectrum tend to be the most stable and undergo less pulsational behavior than neighboring models to the left and right of the peak. At \code{2m\_2h\_25t}, models with M\textsubscript{He} = 50--58\,\Msun\ undergo many weak and some strong pulsations reducing the BH mass and deviating from a linear growth in the BH mass. M\textsubscript{He} = 60--62\,\Msun\ models are the most stable resulting few pulsations, and at M\textsubscript{He} $\geq$\,63\,\Msun, strong pulses mixed with weak pulses result in much less massive BH masses.

All models were run on 8-core nodes and allotted 32GB of RAM. Wall-clock times are generally between 1--4 days for low resolution models (e.g. \code{1m\_1h\_1t} - \code{2m\_2h\_5t}), 1--3 weeks for intermediate resolution models (e.g. \code{5m\_5h\_2p5t} - \code{5m\_2h\_10t}), and 1--10 weeks for the highest resolution models (e.g. \code{2m\_2h\_25t}). The highest resolution models take between 1--3 weeks near the BH mass spectrum peak, between 3--10 weeks to the left of the peak where hundreds of weak pulsations occur, and 3--10 weeks to the right of the peak where strong pulsations are intermixed with weak pulsation episodes. The computational cost of \code{2m\_2h\_25t} limited us from computing full spectrums at higher mass resolutions such as \code{5m\_5h\_25t}. These models could not feasibly be calculated anywhere except at the BH spectrum peak in this work as they would take between 3--9 months to complete. Any model which did not reach CC or result in a PISNe is not included in this work.

In Table \ref{tab:table1} all models except \code{1m\_1h\_t} possess central X(\textsuperscript{12}C)\textsubscript{He} mass fractions within 0.01 of each other at core He-depletion, regardless of the initial He core mass M\textsubscript{He}. The C-O core mass M\textsubscript{CO}\textsubscript{\_He} at He-depletion primarily depends on M\textsubscript{He}. Models with similar initial M\textsubscript{He} develop values of M\textsubscript{CO}\textsubscript{\_He} within 1\,\Msun\ of each other regardless of the adopted resolution. Our CO and He core mass boundaries are defined as the outermost location where X($^{4}$He)$ \leq 0.01$ and X($^{1}$H)$ \leq 0.01$, respectively.

Across all resolutions, the behavior of each PPISN model shown in Figure \ref{fig:fig3} is similar. Regardless of He core mass, each model undergoes He core burning followed by shell He-burning as the core beneath contracts. In all cases, the contracting core is momentarily stabilized by central C-ignition radiatively burning outward and leaving behind an O-Mg rich core, with traces of Si. The next evolutionary phase is primarily determined by the total mass of the C-O core at the end of He-burning which scales linearly with M\textsubscript{He}, and X(\textsuperscript{12}C) of the shell which is identical to X(\textsuperscript{12}C)\textsubscript{He}, the X(\textsuperscript{12}C) mass fraction at He-depletion. If the core is not massive enough or X(\textsuperscript{12}C)\textsubscript{He} is large enough, a convectively burning Carbon shell will form and the core will undergo stable convective O-ignition to maintain hydrostatic equilibrium. This sort of evolution is characterized by a smooth transition from C to O/Ne/Mg and then Si/S/Ar and eventually Fe-burning in the core. This leads the star to CC without any dynamical contractions or pulses driven by $\Gamma_{1}$ instability. These stellar models only lose mass through winds during He-burning, M\textsubscript{eject}$\leq$\,0.3\,\Msun, and thus produce a linear growth in the BH spectrum in correspondence with the initial He core mass. For $\sigma$[\textsuperscript{12}C($\alpha$,$\gamma$)\textsuperscript{16}O]$=0$ the left side of the BH mass spectrum is stable up to M\textsubscript{He} $\simeq$\,47\,\Msun, at which point the lower edge of the pulsational pair-instability strip is encountered. Models that are slightly more massive than M\textsubscript{He} $\simeq$\,47\,\Msun\ and or with slightly lower X(\textsuperscript{12}C)\textsubscript{He} begin to suffer from pair-instability in their interiors fueling a dynamic evolution toward CC. Just above M\textsubscript{He} $\simeq$\,47\,\Msun, pair-production in the interior of these stellar cores softens the equation of state, and dynamical contractions begin to drive weak pulses that remove material from the surface of these models. As M\textsubscript{He} increases, weak pulses increase in frequency and number until strong pulses develop, unbinding M\textsubscript{eject}$\geq1$\,\Msun until the peak of the BH mass spectrum is encountered. More detail on the characteristics of weak pulsations encountered before the peak have been discussed in \citet{woosley_2017_aa,Renzo_2020b_aa}, although not at the highest resolutions explored in this work. Weak pulsations begin to develop when convective shell C-burning can not supply the energy needed to stabilize the shell long enough for smooth O-ignition in the center. The onset of this instability has been discussed in detail in the behavior of the 50\,\Msun\ He cores from section \ref{s.mixing_floors}. stronger pulsations near to and at the peak of the BH mass spectrum typically involve a short episode of convective shell C-burning while a portion of the central O-core convectively burns. Once $^{12}$C is nearly depleted from the shell, $\Gamma_{1}$ instability leads to a dynamical contraction followed by a thermonuclear explosion of a significant mass fraction of the O-rich core. When the shock reaches the surface of the model, multiple contractions and bounces occur as the star relaxes back to hydrostatic equilibrium. After a dynamical episode, stable O-burning can ensue in the core, followed by burning of heavier isotopes and eventually CC.

To the right of the BH mass spectrum peak, all models with larger M\textsubscript{He} are unable to develop sufficient pressure from shell C-burning to enable smooth O-ignition in the core. The convective growth of the C-burning shell is undercut by a $\Gamma_{1}$ instability resulting in an infalling core that radiatively burns through most of its neon on a timescale of $\lesssim50-70$ seconds followed by explosive burning of the central O-core in the next $\lesssim 50$ seconds. Nearly all of the central $^{16}$O is burned in this explosive burning episode, removing between $\simeq$\,5--30\,\Msun\ of material after just one strong pulse and the weak pulses that follow. These pulses are so strong they temporarily perturb the model for $1-10^{3}$ years. After the model relaxes back into hydrostatic equilibrium, nuclear burning recommences and one or several more strong and weak pulsing episodes may follow depending on the strength of the initial pulse and the amount of unburnt $^{16}$O remaining in the core. For models with M\textsubscript{He}$\geq$\,72\,\Msun, the O-rich core explodes in a single thermonuclear pulse driving a supersonic shock with energy greater than the  total binding energy of the model, disrupting the entire model after a single pulse, and leaving no compact remnant in a PISNe.

In panel 1 of Figure \ref{fig:fig3} we compare the BH mass spectrum at three resolutions with increasing mass resolution and otherwise equal temporal resolution. Models with M\textsubscript{He}$\leq$\,47\,\Msun\ do not appear to encounter pulsational pair-instability according to panel 1, as this appears to be the lower edge of the pair-instability strip. The first resolution to display PPISN behavior is \code{5m\_5h\_2p5t}, followed by \code{5m\_2h\_2p5t}, \code{2m\_2h\_2p5t}, and then \code{1m\_1h\_2p5t}. Increasing the mass resolution appears to smooth out the peak of the BH mass Spectrum, however this could be due to a lack of temporal resolution to accurately resolve models with greater mass resolution. The erratic behavior near the peak for \code{1m\_1h\_2p5t} is likely due to being slightly under-resolved. Convergence in the peak BH mass is observed between \code{5m\_5h\_2p5t} and \code{5m\_2h\_2p5t}. Both resolutions predict a BH mass peak of M\textsubscript{BH}$\simeq$\,54.8\,\Msun\ satisfying our criteria for convergence at the $<$\,1\,\Msun\ level. This convergence is slightly suspect though, as both models evolved from a 55\,\Msun\ He core that does not enter the hydrodynamic solver during core O-burning, remaining in the implicit hydrostatic solver until the HLLC hyrodynamic solver is turned on at CC, refer to \citet{marchant_2019_aa} for a discussion of the criterion to turn on the HLLC solver. Directly right of the peak, all the model resolutions appear to fall within 1--2\,\Msun\ of one another.

In panel 2 of Figure \ref{fig:fig3} we compare our previously converged mass resolution models with two additional resolutions at twice the temporal resolution with and without a minimum diffusive mixing floor, \code{5m\_5h\_5t} and \code{5m\_5h\_5t\_D}. This comparison illustrates that while the mass resolution might be converged, increasing the temporal resolution and including a minimum diffusive mixing floor changes the peak BH mass at $>$1\,\Msun\ level indicating our models are not fully time resolved near the peak. There is a wide spread in BH mass across the entire peak, another indication that increasing the temporal resolution of these models might be warranted.

In panel 3 we conduct a test of the temporal resolution of these models by controlling for changes in the central density per timestep. Increasing from \code{2m\_2h\_2p5t}  to \code{2m\_2h\_25t}, an order of magnitude increase in temporal resolution, leads to a change in the BH mass of $\simeq$\,8\Msun. This is a significant increase in the peak BH mass demonstrating the importance of time resolution in resolving the convective burning behavior of our PPISN models.  To the left of the peak, the number of weak pulses resolved is also particularly sensitive to time resolution. This is one of the primary reasons \code{2m\_2h\_25t} takes $>$1,000,000 time steps in models between 50--58\,\Msun. These models are able to resolve hundreds of weak pulses creating small shocks that unbind between 0.5--1\,\Msun\ of material. In several cases (e.g. M\textsubscript{He}$= 53, 54, 55, 59$\,\Msun) these weak shocks precede the development of an episode of strong shocks which can go entirely unresolved at resolutions lower than \code{2m\_2h\_25t}, straying from the trend of linear growth in M\textsubscript{BH} observed at lower resolutions. Hundreds of weak shocks are also observed on the right side of the BH mass spectrum peak. Between each strong shock hundreds of weak shocks can remove multiple \Msun\ of additional material.

In panel 4, we show the resolved peak of the BH spectrum from our highest temporal resolution models and confirm its existence at our highest adopted mass resolution. By comparing our highest temporal resolution \code{2m\_2h\_25t} with \code{5m\_2h\_10t} and \code{5m\_2h\_10t\_D} we confirm that 10x the temporal resolution is needed to resolve the BH mass spectrum peak to within $\leq$\,1\,\Msun. We confirm the convergence of these models by including the peak BH masses from \code{5m\_2h\_25t}, \code{5m\_5h\_25t}, \code{5m\_5h\_25t\_D}. In conjunction with Table \ref{tab:table1}, Panel 4 illustrates that the adoption of a minimum diffusion coefficient $D_{\rm min}$ changes the peak BH mass at the $\leq$\,0.1\,\Msun\ level. Models with $D_{\rm min}$ also tend to be more numerically stable than models run without, encountering less overall crashes, and typically achieving their final fate in less timesteps and therefore a generally lower wall-clock time, even for similar BH masses.

Looking at all four panels, we are able to differentiate between the importance of mass and temporal resolution. The amount of mass resolution needed to resolve the peak is limited by the amount of temporal resolution available. For example, models run with \code{2m\_2h\_5t} are more stable than models run with \code{5m\_5h\_5t} displaying respective peak BH masses of M\textsubscript{BH}$\simeq$\,59.1\,\Msun\ and M\textsubscript{BH}$\simeq$\,53.4\,\Msun. We find that an adequate amount of temporal resolution is necessary to resolve the peak at a given mass resolution. Models with large mass and low time resolution tend to produce the most smeared peaks with the lowest overall peak BH mass, where as models with the low mass resolution and high temporal resolution display  the largest overall peak BH mass. This explains why \code{1m\_1h\_2p5t} is able to resolve a larger peak BH mass than at any other mass resolution with similar time resolution. Increasing mass resolution in a vacuum is not advised, without comparable increases in temporal resolution. Panel 4 indicates that time steps limited by at least \code{$\delta$\textsubscript{log$\rho_{c}$}}$\leq 2.5\times10^{-4}$ are needed to resolve the peak BH mass, and \code{$\delta$\textsubscript{log$\rho_{c}$}}$\leq 10^{-4}$ is needed to accurately resolve the shape of the BH mass spectrum, while the amount of mass resolution needed is rather unclear. 

\citet{marchant_2020_aa} and \citet{woosley_2021_aa} found that the efficiency of angular-momentum transport
changes the lower edge of the BH's mass gap at the $\simeq$\,10\% level.
\citet{farmer_2019_aa} found that the lower edge of mass gap was robust at the $\simeq$\,10\% level
to changes in the metallicity, wind mass loss prescription, and treatment of chemical mixing. In \citet{2020_renzo_ras} on convection during pulses, we showed that the treatment of time-dependent convection does not directly affect the maximum BH mass. In \citet{mehta_2022_aa}, we claimed our models for the peak BH mass were robust with respect to mass and temporal resolution at the $\simeq$\,10\% level. In this section we have gone further to show that by adopting a resolution of \code{5m\_2h\_10t} or \code{5m\_2h\_10t\_D} our peak BH mass models are robust to mass and temporal resolution at the $\simeq$\,1\% level at $\sigma$[$^{12}$C($\alpha$,$\gamma$)$^{16}$O] = $0$.

\begin{figure}[!htb]
    \centering
    \includegraphics[width=3.38in]{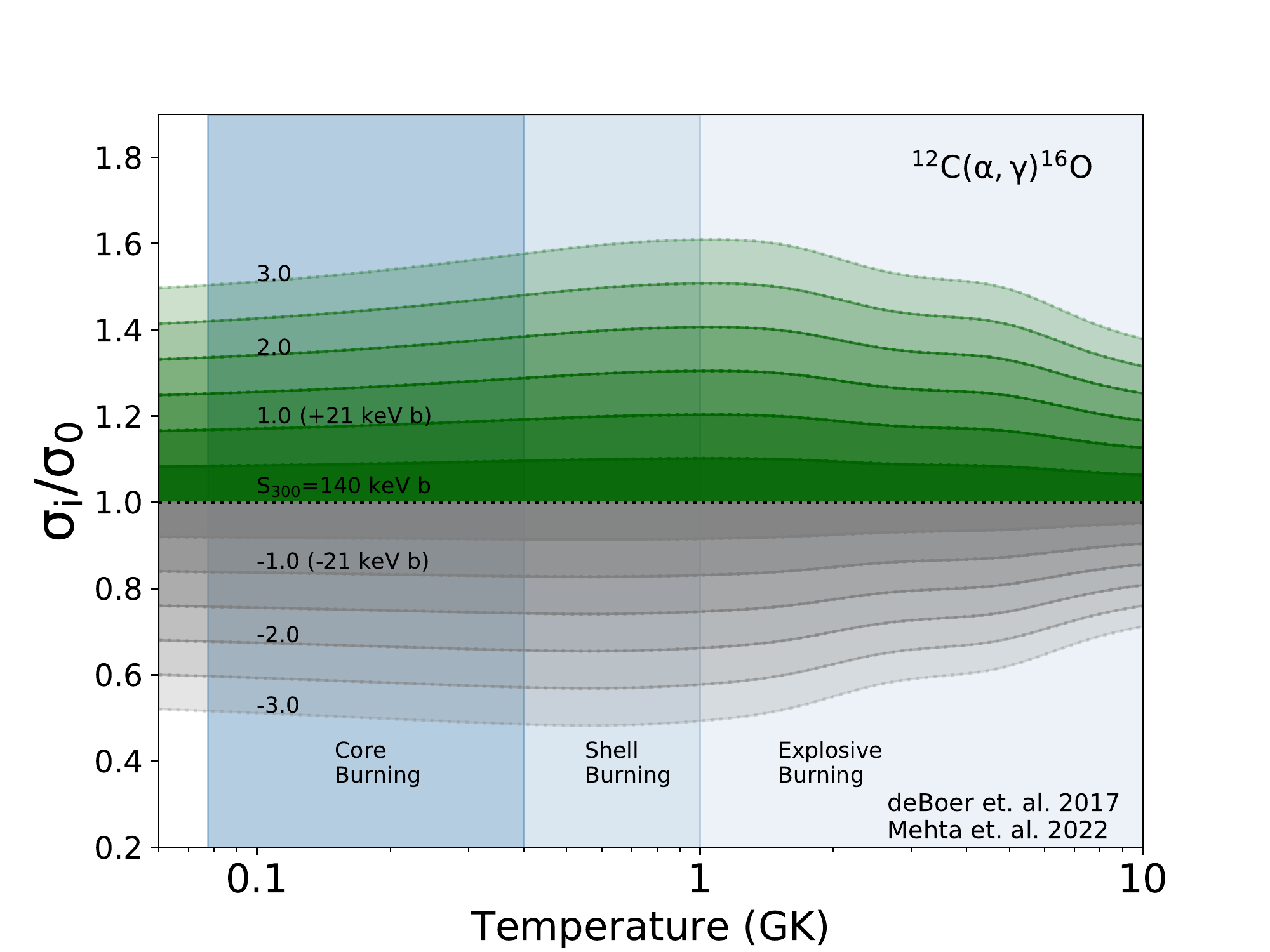}
    \caption{ $^{12}$C($\alpha,\gamma$) reaction rate ratios, $\sigma_i/\sigma_0$, as a function of temperature.
$\sigma_i$ spans -3.0 to 3.0 in 0.5 step increments, with $\sigma_0$ being the current nominal rate.  Negative $\sigma_i$ are gray curves and positive $\sigma_i$ are green curves. The $\pm$\,1,\,2,\,3 $\sigma_i$ curves are labeled.  The blue band show the range of temperatures encountered during core and shell He burning.
}
    \label{fig:fig4}
\end{figure}

\section{The Rate Dependent Single Star BH Mass Gap}\label{s.bh_massgap}

During the core He-burning phase of stellar evolution, a competition
between the (3-$\alpha$) process $^{4}$He$(\alpha,)^{8}$Be$(\alpha,)^{12}$C$^{*}(,2\gamma)^{12}$C and
$^{12}$C($\alpha$,$\gamma$)$^{16}$O establishes the C/O ratio of the
stellar core for later C and O burning phases. The evolution of He
cores through advanced burning stages depends sensitively on the C/O
ratio of these models at He-depletion. Reducing the uncertainty in the $^{12}$C($\alpha$,$\gamma$)$^{16}$O reaction rate probability distribution function
is a goal of forthcoming experiments \citep{deboer_2017_aa,smith_2021_aa}. 
A common approach to
investigating the C/O ratio of differing stellar cores is to explore a
modified $^{12}$C($\alpha$,$\gamma$)$^{16}$O during He-burning, and
tracing it's effect on the BH mass distribution \citep{takahashi_2018_aa,farmer_2020_aa,costa_2021_aa,mehta_2022_aa}. In
this section, we will continue with a similar exploration of this rate
albeit at a significantly higher model resolution than similar works
previously conducted in \MESA. We will explore variations in the
$^{12}$C($\alpha$,$\gamma$)$^{16}$O at the 0.5$\sigma$ level, with the
$^{12}$C($\alpha$,$\gamma$)$^{16}$O rate provided by
\citet{deboer_2017_aa} and updated in \citet{mehta_2022_aa}, see
Figure \ref{fig:fig4}. $\sigma_{o}$ represents the median rate
consistent with an astrophysical S-factor of S(300 keV) = 140 keV b
with a $\pm 1 \sigma = 21$ keV b uncertainty. By exploring $\pm
3\sigma$, we effectively explore the range S(300 keV) = (77,203) keV
b, where positive and negative $\sigma$ indicate a stronger and weaker
rate than the median value, respectively.

In this Section,  we will explore a range of \MESA\ stellar model
resolutions across the rate dependant upper mass gap. In section
\ref{s.c12ag_mass_gap} we extend our resolution study to resolve
the tip of the BH mass spectrum at the $\leq$\,2\,\Msun\ level at $\sigma$[$^{12}$C($\alpha$,$\gamma$)$^{16}$O] = $-3,+3$ and
we recompute a new resolved lower edge to the BH mass gap. In section
\ref{s.triple_alpha_bh_massgap}, we discuss the impact of a newly
calculated $^{4}$He$(2\alpha,)^{12}$C (3-$\alpha$) reaction rate on
the location of the lower edge of the
$^{12}$C($\alpha$,$\gamma$)$^{16}$O rate-dependant mass gap. We then
recompute a resolved lower edge to the rate-dependent mass gap with
inclusion of this revised 3-$\alpha$ rate. We end this section by
comparing our new lower BH mass gap edge with recent measurements of
binary black holes (BBH) mergers detected in the third Gravitational-Wave Transient Catalogue (GWTC-3) by
\citep{the-ligo-scientific-collaboration_2021_aa}.

\subsection{Resolving the $^{12}$C($\alpha$,$\gamma$)$^{16}$O Dependant BH Mass Gap}\label{s.c12ag_mass_gap}

\begin{figure}
    \centering
    \includegraphics[width=3.38in]{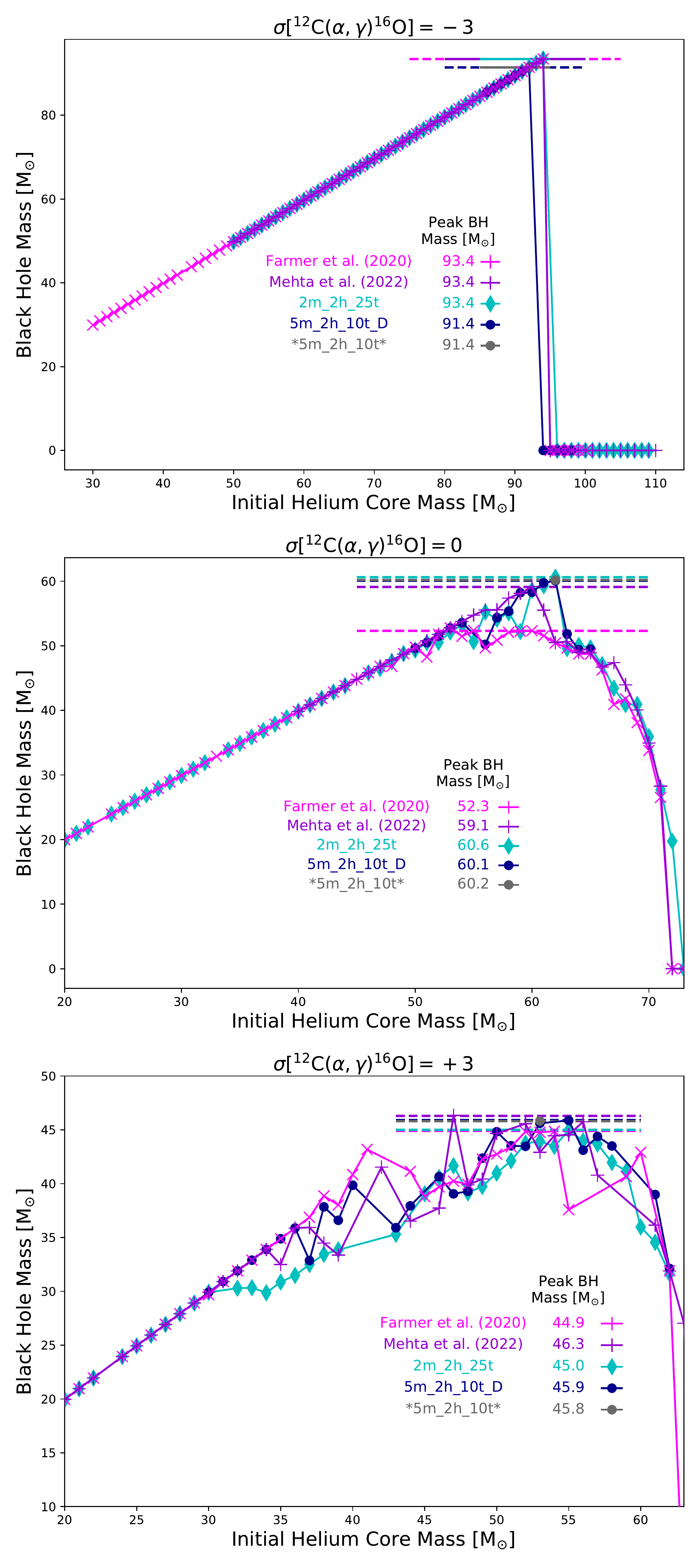}
    \caption{ The BH mass spectrum for different values of the $^{12}$C($\alpha$,$\gamma$)$^{16}$O rate; left ($\sigma = -3$), right ($\sigma = +3$), and center ($\sigma = 0$), for five of the resolutions listed in Table \ref{tab:table1}. All panels share a similar x-axis, Initial He Core Mass M\textsubscript{He} in \Msun, and y axis, BH Mass in \Msun. Resolutions that are only shown at their respective peak BH mass are labeled with asterisks on either side. Horizontal dashed and solid lines indicate the location of the peak BH mass for each resolution.}
    \label{fig:fig5}
\end{figure}

\begin{figure*}
    \centering
    \includegraphics[width=5in]{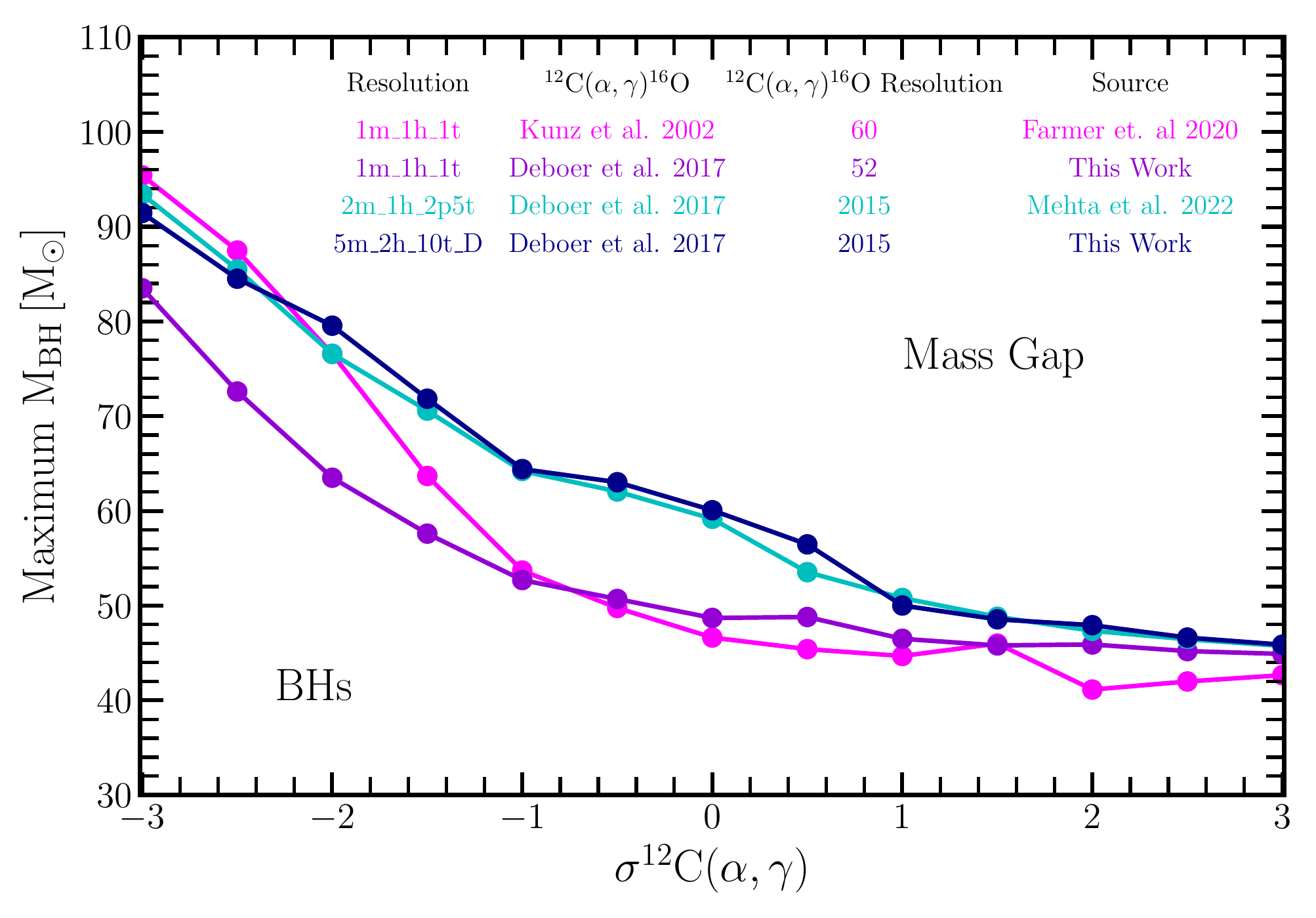}
    \caption{The location of the lower edge of the BH mass gap as a function of the
temperature-dependent uncertainty in
the \textsuperscript{12}C($\alpha$,$\gamma$)\textsuperscript{16}O
reaction rate. Each line marks the lower mass-gap boundary predicted by
the adopted \textsuperscript{12}C($\alpha$,$\gamma$)\textsuperscript{16}O rate
uncertainties. The magenta/pink line mark the lower mass-gap boundary,
as found in Figure~5 of \citet{farmer_2020_aa}, predicted by the
\citet{kunz_2002_aa} rate as expressed in the \texttt{STARLIB}
reaction-rate library \citep{sallaska_2013_aa}. The dark purple line marks the lower mass-gap boundary predicted by models adopting the resolution from \citet{farmer_2020_aa} and the original 52 temperature point \textsuperscript{12}C($\alpha$,$\gamma$)\textsuperscript{16}O rate provided by \citet{deboer_2017_aa}. The coral blue line marks the lower mass-gap boundary using the updated \textsuperscript{12}C($\alpha$,$\gamma$)\textsuperscript{16}O calculated and implemented in \citet{mehta_2022_aa}. The dark blue line marks the revised lower mass-gap boundary calculated in this work using the updated \textsuperscript{12}C($\alpha$,$\gamma$)\textsuperscript{16}O adopted in \citet{mehta_2022_aa}. Note that $\sigma$[$^{12}$C($\alpha$,$\gamma$)$^{16}$O] represent different astrophysical S-factor's for each rate source.}
    \label{fig:fig6}
\end{figure*}

By varying the $^{12}$C($\alpha$,$\gamma$)$^{16}$O rate, we probe the convergence of models with differing C/O cores. We then establish a new resolved lower edge of the upper mass gap as M\textsubscript{lower} $\simeq$\,60$^{+32}_{-14}$\,\Msun\ from the $\pm 3\sigma$ uncertainty in the $^{12}\text{C}(\alpha, \gamma) ^{16}\text{O}$. In Figure \ref{fig:fig5} we compute the BH mass spectrum and their associated peak BH masses at five different resolutions found in Table~\ref{tab:table1}. In each panel we display the BH mass spectrums produced by adopting the model resolutions found in \citet{farmer_2020_aa} (\code{1m\_1h\_1t}), \citet{mehta_2022_aa} (\code{2m\_1h\_2p5t}), our highest feasible temporal resolution \code{2m\_2h\_25t}, and \code{5m\_2h\_10t\_D} which we've shown to reproduce the BH mass spectrum peak to within $\leq$\,1\,\Msun\ at $\sigma$[$^{12}$C($\alpha$,$\gamma$)$^{16}$O] = $0$. We opt to only compute BH mass spectrums across the peak for \code{5m\_2h\_10t\_D} as these models tend to be much more numerically stable than models run with \code{5m\_2h\_10t}, encountering less overall crashes than models which do not adopt $D_{\rm min}$. In each panel only the peak BH mass is shown for \code{5m\_2h\_10t} to illustrate similar agreement with \code{5m\_2h\_10t\_D}.

In the top panel of Figure \ref{fig:fig5}, we show the BH mass spectrum for models with $\sigma$[$^{12}$C($\alpha$,$\gamma$)$^{16}$O] = $-3$. These models possess larger C/O mass fractions with typical X(\textsuperscript{12}C)$\sim 0.32$. The larger $^{12}$C mass fraction of these He cores allows for the development of strong convective shell C-burning which manages to counter the affects of pair-instability in these cores long enough for them to reach CC \citep{takahashi_2018_aa,woosley_2021_aa}. There is little shape to this spectrum as models are stable up until they undergo PISNe. If the true cosmic $^{12}$C($\alpha$,$\gamma$)$^{16}$O rate was in fact this weak, we would not expect to observe any PPISN stars as the lower edge of the pair instability strip is synchronous with the lower edge of the BH mass gap. Here, \citet{farmer_2020_aa} (\code{1m\_1h\_1t}), \citet{mehta_2022_aa} (\code{2m\_1h\_2p5t}), and \code{2m\_2h\_25t} produce BHs of identical mass, M\textsubscript{BH} $\simeq$\,93.4\,\Msun. Models \code{5m\_2h\_10t} and \code{5m\_2h\_10t\_D} appear to agree with one another, producing BHs with identical masses M\textsubscript{BH} $\simeq$\,91.4\,\Msun. Interestingly our lowest resolution illustrates better agreement with out highest resolution than our moderate resolution models. As a check, we have computed an additional model with \code{5m\_2h\_25t} and found the peak BH mass remains M\textsubscript{BH} $\simeq$\,93.4\,\Msun. It is unclear why our lowest resolution models agree with our highest resolution models here, however in a stellar evolution model there are many choices that are made which interact highly non-linearly with one another. It is possible that our experiment varying only a few parameters does not capture all the possible variations necessary. Nonetheless, the location of the BH mass spectrum peak is sensitive to temporal resolution due to its discontinuous nature. In this case, we are able to resolve the peak BH mass spectrum at lower resolutions, but we can only be confident in our peak BH mass estimate to within $\leq$\,2\,\Msun\ with \code{5m\_2h\_10t\_D}.

In the middle panel of Figure \ref{fig:fig5}, we show the BH mass spectrum for models with $\sigma$[$^{12}$C($\alpha$,$\gamma$)$^{16}$O] = $0$. The behavior of models with these C/O mass fractions, X(\textsuperscript{12}C)$\sim 0.17$, were previously discussed in section \ref{s.bh_res_spectrum}. Within the context of our \MESA\ models,
\citet{mehta_2022_aa} has shown the dependence of the BH mass spectrum
on the tabulated temperature resolution of the $\sigma$\,=0 $^{12}$C($\alpha$,$\gamma$)$^{16}$O reaction rate
at \code{2m\_1h\_2p5t}. When the reaction rate was defined by 52 temperature points, the BH mass spectrum reaches a
maximum BH mass of 49.6\,\Msun\ at an initial He core mass of 55.0\,\Msun.
When the reaction rate is defined by 2015 temperature points, the BH mass spectrum reaches a
maximum BH mass of 59.1\,\Msun\ at an initial He core mass of 60.0\,\Msun. In this work, we have confirmed the maximum BH mass to be 60.3--60.4\,\Msun\ with an initial He core mass of 62\,\Msun\ at our highest resolution, \code{5m\_5h\_25t}. \citet{mehta_2022_aa} happened to choose a combination of mass and temporal resolution to be able to resolve the peak of the BH mass to within 2\,\Msun\  of the resolved value. As a result, the ability of the 2015 point rate to produce models which sustain a linear trend of larger BH masses with larger initial He core masses was not just a result of the improved tabulated resolution of the $^{12}$C($\alpha$,$\gamma$)$^{16}$O reaction rate, but also largely due to the resolution of the models used to calculate the peak in \citet{mehta_2022_aa}. In this work, the peak of the $\sigma$[$^{12}$C($\alpha$,$\gamma$)$^{16}$O] = $0$ BH mass spectrum can only be resolved to within $\leq$\,1\,\Msun\  with \code{5m\_2h\_10t\_D} or greater resolutions.

In the bottom panel of Figure \ref{fig:fig5}, we show the BH spectrum for models with $\sigma$[$^{12}$C($\alpha$,$\gamma$)$^{16}$O] = $+3$. These models possess the lowest overall C/O mass fractions with typical X(\textsuperscript{12}C)$\sim 0.07$. Given the small C/O mass fractions present in these cores, most of these models undergo at least one strong pulse. At the lower end of the BH mass spectrum a notable difference in the location of the lower edge of the pair-instability strip can be seen between \code{2m\_2h\_25t} and all other resolutions. From this panel, we find that the lower edge of the pair instability strip is $\simeq$\,32\,\Msun\ at \code{2m\_2h\_25t}, $\simeq$\,5\,\Msun\ lower than at any other resolution. Temporal resolution appears to be of critical importance in establishing the lower edge of the pair-instability strip for models with low C/O mass fractions. Models with low temporal resolution are unable to resolve the hundreds of weak pulses which  which culminate in a strong pulse for models with M\textsubscript{He} as low as 32\,\Msun. Across the mass spectrum, a large spread in the BH masses is found at varying resolutions with the primary difference being owed to the shear number of weak pulsation episodes intermixed between strong pulses. Models run at \code{2m\_2h\_25t} are able to resolve many more weak pulses than at any other resolution, leading to BH masses which are typically lower than those predicted by lower resolutions. At $\sigma$[$^{12}$C($\alpha$,$\gamma$)$^{16}$O] = $+3$, we are able to resolve the peak BH mass to within $\leq$\,1\,\Msun\ with \code{5m\_2h\_10t\_D}, although we cannot resolve the entire shape of the BH mass spectrum.

Figure \ref{fig:fig5} shows that the temporal resolution during the hydrodynamic phase of evolution is most important for resolving the peak of the BH mass spectrum. 
This is due to a tight coupling between the nuclear burning and convection at each timestep.
Models with $\sigma$[$^{12}$C($\alpha$,$\gamma$)$^{16}$O]\,=\,$-$3 spend the least amount of time in the hydrodynamic phase, while models with $\sigma$[$^{12}$C($\alpha$,$\gamma$)$^{16}$O]\,=\,+3 spend the most. The BH mass spread between different temporal resolutions is larger the longer these models spend in the hydrodynamic phase. Models with stronger $^{12}$C($\alpha$,$\gamma$)$^{16}$O rates undergo stronger nuclear burning episodes and convection, requiring the highest temporal resolutions to resolve.

\begin{figure*}
    \centering
    \includegraphics[width=5in]{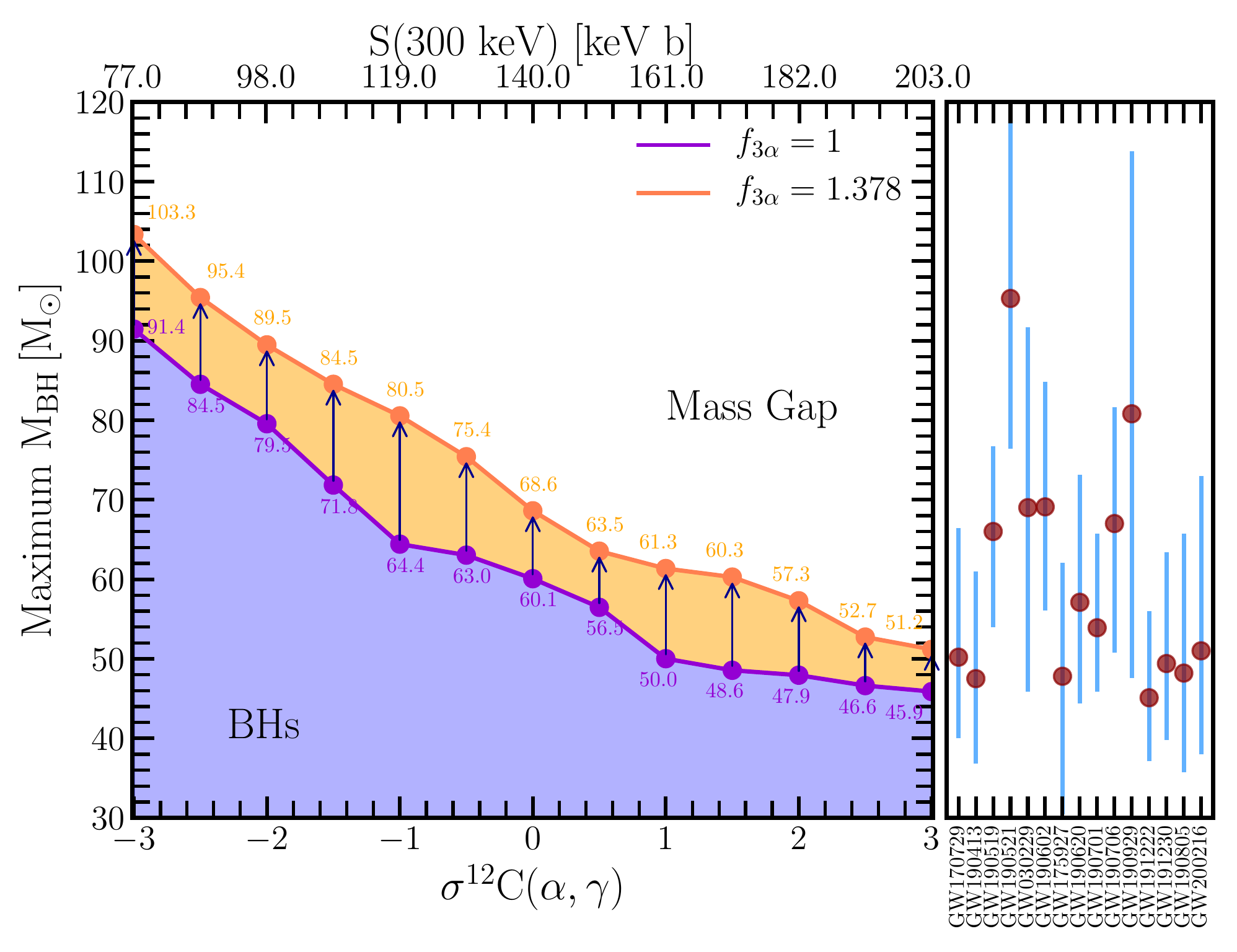}
    \caption{The location of the lower edge of the BH mass gap as a function of the
temperature-dependent uncertainty in
the \textsuperscript{12}C($\alpha$,$\gamma$)\textsuperscript{16}O
reaction rate. The dark purple line marks the revised lower mass-gap boundary calculated in this work using the updated \textsuperscript{12}C($\alpha$,$\gamma$)\textsuperscript{16}O adopted in \citet{mehta_2022_aa} and $f_{3\alpha}=1$ \citep{angulo_1999_aa}. The orange line marks the revised lower mass-gap boundary calculated in this work using the updated \textsuperscript{12}C($\alpha$,$\gamma$)\textsuperscript{16}O adopted in \citet{mehta_2022_aa} and $f_{3\alpha}=1.378$ \citep{kibedi_2020_prl}. The white region denotes the mass gap and the
yellow-orange regions highlight differences between models computed with $f_{3\alpha}$\,=\,1 and $f_{3\alpha}$\,=\,1.378 at the same  uncertainty in \textsuperscript{12}C($\alpha$,$\gamma$)\textsuperscript{16}O. All BH masses with M\textsubscript{BH}$\geq$\,45\,\Msun\ included in GWTC-3 are shown with dark red points and blue error bars showing the uncertainty in their inferred mass. The approximate astrophysical S-factor corresponding to each probability distribution function sourced $\sigma$ value is labeled on the top x axis.}
    \label{fig:fig7}
\end{figure*}

Given the large computational cost of computing models with \code{2m\_2h\_25t}, we adopt \code{5m\_2h\_10t\_D} as our highest resolution moving forward, with numerical convergence in peak BH mass at the $\leq$\,(2, 1, 1)\,\Msun\ level for $\sigma$[$^{12}$C($\alpha$,$\gamma$)$^{16}$O] = $-3, 0, +3$. More specifically, models run with \code{5m\_2h\_10t\_D} appear to underestimate the peak BH mass by $\leq$\,2\,\Msun\ at $\sigma = -3$ and overestimate the peak BH mass by $\leq$\,1\,\Msun\ at $\sigma = 3$. In Figure \ref{fig:fig6} we compute the lower edge of a resolved $^{12}$C($\alpha$,$\gamma$)$^{16}$O rate-dependant BH mass gap and compare to previous work. In \citet{farmer_2020_aa}, using a default resolution of \code{1m\_1h\_1t}, a 60 temperature point $^{12}$C($\alpha$,$\gamma$)$^{16}$O from STARLIB \citep{kunz_2002_aa} was calculated and discussed along with the \citet{deboer_2017_aa} $^{12}$C($\alpha$,$\gamma$)$^{16}$O rate at $\sigma=-1,0,1$. In this work we extend that calculation across the lower edge of the mass gap to show the result of adopting the original 52 temperature point \citet{deboer_2017_aa} $^{12}$C($\alpha$,$\gamma$)$^{16}$O rate at \code{1m\_1h\_1t}. While the agreement between  \citep{deboer_2017_aa} and \citep{kunz_2002_aa} was $\leq$\,3\,\Msun\ at $\sigma=-1,0,1$, the difference between both rates grows appreciably large between $\sigma=-1,-3$. This is due to the fact that $\sigma$ does not represent the same physical quantity, the astrophysical S-factor, for different sources of the reaction rate.

In \citet{mehta_2022_aa}, a revised \citet{deboer_2017_aa} reaction rate probability distribution function defined by 2015 temperature points at each $\sigma_i$ was adopted and implemented at an improved resolution of \code{2m\_1h\_2p5t}. This small improvement in model and reaction rate resolution substantially increased the lower edge of the BH mass gap between  $\sigma=-2,+3$. The resolved lower edge computed with \code{5m\_2h\_10t} shows reasonable agreement with the results of \citet{mehta_2022_aa} at the $\leq$\,4\,\Msun\ level across the lower edge of the BH mass gap.

\begin{deluxetable*}{cccclc}[!htb]
  \tablenum{2}
  \tablecolumns{6}
  \tablewidth{\apjcolwidth}
  \tablecaption{\mesa\ BH mass gap He core explosive behavior for Figure \ref{fig:fig7}}\label{tab:table2}
  \tablehead{
    \colhead{$\sigma ^{12}$C$(\alpha,\gamma)$} & \colhead{\code{M\textsubscript{He}} (\Msun)}& \colhead{\code{M\textsubscript{CO}} (\Msun)} & \colhead{\code{{X(\textsuperscript{12}C)}}} & \colhead{\code{Pulses}}&
     \colhead{\code{M\textsubscript{BH}} (\Msun)}    }
  \startdata
      \code{$f_{3\alpha}=1$} & & & & & \\
      -3.0& 92 & 82.80 & 0.3156 & 0 & 91.45\\
      -2.5& 85 & 76.57 & 0.2816 & 1 weak & 84.51    \\
      -2.0& 80 & 72.12 & 0.2510 & 0 & 79.55    \\
      -1.5& 73 & 72.61 & 0.2278 & 1 strong, $\sim10$ weak &  71.84   \\
      -1.0& 65 & 58.93 & 0.2105 & $>100$ weak &  64.41   \\ 
      -0.5& 65 & 59.03 & 0.1855 & 1 strong, $\sim10$ weak &  63.03   \\
       0  & 62 & 56.39 & 0.1673 &  1 strong, $\sim10$ weak & 60.07  \\
       0.5& 58 & 52.37 & 0.1508 & 1 strong, $\sim9$ weak &  56.47   \\
       1.0& 56 & 49.73 & 0.1280 & 4 strong, $>100$ weak & 50.01    \\ 
       1.5& 56 & 50.10 & 0.1160 & 3 strong, $>80$ weak &  48.55   \\
       2.0& 56 & 49.69 & 0.0972 & 3 strong, $>30$ weak & 47.94    \\
       2.5& 55 & 49.07 & 0.0840 & 3 strong, $>30$ weak & 46.63    \\
       3.0& 55 & 48.89 & 0.0729 & 3 strong, $>20$ weak &  45.88   \\
      \hline{}
      \code{$f_{3\alpha}=1.378$} &   &  &  &   \\
      -3.0& 104 & 93.70 & 0.3689 & 0 & 103.34   \\
      -2.5& 96  & 86.49 & 0.3356 & 0 & 95.41    \\
      -2.0& 90  & 81.07 & 0.3061 & 0 & 89.47   \\
      -1.5& 85  & 76.58 & 0.2800 & 0 &  84.51   \\
      -1.0& 81  & 73.04 & 0.2564 & 0 & 80.54    \\
      -0.5& 76  & 68.60 & 0.2370 & $>90$ weak &  75.41  \\
       0& 70    & 63.27 & 0.2216 & 1 strong, $\sim9$ weak &  68.61   \\
       0.5& 65  & 59.03 & 0.2057 & 1 strong, $\sim9$ weak &  63.53   \\
       1.0& 63  & 57.31 & 0.1877 & 1 strong, $\sim9$ weak & 61.35   \\
       1.5& 62  & 56.42 & 0.1700 & 1 strong, $\sim10$ weak &  60.27   \\
       2.0& 60  & 54.57 & 0.1553 & 1 strong, $>100$ weak &  57.29   \\
       2.5& 57  & 51.70 & 0.1437 & 2 strong, $>20$ weak & 52.73    \\
       3.0& 55  & 48.72 & 0.1274 & 2 strong, $>100$ weak & 51.18    \\
  \enddata
 \tablenotetext{}{ All models are computed with \code{5m\_2h\_10t}. M\textsubscript{CO} and X(\textsuperscript{12}C) are respectively the C-O core mass and central $^{12}$C mass fraction at core He-depletion.
 }
\end{deluxetable*}

\subsection{The Impact of 3-$\alpha$ on the BH Mass Gap and GWTC-3}\label{s.triple_alpha_bh_massgap}

 The 3-$\alpha$ process is primarily carried out by the fusion of three $\alpha$ particles to form \textsuperscript{12}C in the Hoyle-state \citep{hoyle_1954_aa}, a resonant 0\textsuperscript{+} second excited state of \textsuperscript{12}C at $\sim$7.65 Mev, which then decays into the ground state. The currently adopted rate for 3-$\alpha$ in many stellar evolution codes currently comes from \citet{nomoto_1985_ab} which adopted $\Gamma$\textsubscript{rad}$= 3.7$ meV (i.e., 3.7$\times$10$^{-3}$ ev) as the radiative branching ratio width of the Hoyle-state. All three 3$\alpha$ rates available for use in \MESA~adopt this value of $\Gamma$\textsubscript{rad} in their calculation: CF88 \citep{caughlan_1988_aa}, NACRE \citep{angulo_1999_aa}, and JINA REACLIB \citep{cyburt_2010_ab}. We have up to this point adopted the NACRE formulation for 3-$\alpha$ rate though it should be noted that the CF88 and JINA rate are always smaller than the NACRE rate at He-burning temperatures (0.2-0.4 GK), see Figure \ref{fig:fig8} in Appendix A. More recently \citet{Freer_2014_ppnp} have performed an up to date review of the relevant literature to affirm the recommended value to as $\Gamma$\textsubscript{rad}$= 3.7$ meV. Recent measurements of by \citet{2020_Eriksen_pr,kibedi_2020_prl,2021_Cook_prc} have measured and proposed a revised width of $\Gamma$\textsubscript{rad}$= 5.1(6)$ meV. If true, this would imply an upward revision in the 3$\alpha$ rate at the 37.8\% level. An increase of this magnitude would warrant a large revision to the BH mass gap and could improve the overall agreement between recent gravitational wave detection of BBHs and the theorized existence of pair-instability stars. In \citet{1988_rolfs_book} the 3-$\alpha$ reaction rate is quoted as being known to a 15\% accuracy, and to a 10\% accuracy in \citet{west_2013_aa} and \citet{2014_austin_prl}.

We explore the impact of the 3-$\alpha$ reaction rate by implementing a 37.8\% larger rate as a multiplicative factor on the currently adopted NACRE rate. Here, the NACRE rate is represented by $f_{3\alpha}$\,=\,1, while the revised rate denoted by $f_{3\alpha}$\,=\,1.378. Figure \ref{fig:fig7} shows the location of the lower edge to the $^{12}$C($\alpha$,$\gamma$)$^{16}$O rate-dependant BH mass gap at a resolution of \code{5m\_2h\_10t\_D} for $f_{3\alpha}$\,=\,(1, 1.378), supplemented with the data provided in Table \ref{tab:table2}. Figure \ref{fig:fig7} also shows all BHs that with M $\geq$\,45\,\Msun\ in GWTC-3 \citep{abbott_2022_aa}. 
Testing at a resolution of \code{5m\_2h\_25t} for 
$\sigma$\,=\,$-$3 confirms that the peak BH mass of M\textsubscript{BH}\,=\,103.3\,\Msun\ for $f_{3\alpha}$\,=\,1.378
remains within $\leq$\,2\,\Msun\ of M\textsubscript{BH}\,=\,105.3\,\Msun, the value found by models calculated with \code{5m\_2h\_25t}. 
The BH mass gaps shown in Figure \ref{fig:fig7}
are likely to be numerically resolved with $\lesssim$\,2\,\Msun\ as compared to models 
run at our highest temporal resolution. At
$\sigma$[$^{12}$C($\alpha$,$\gamma$)$^{16}$O] = $-3, 0, +3$, using
$f_{3\alpha}$\,=\,1.378 results in an increase in the lower edge of the
upper stellar mass gap of M\textsubscript{BH} = (11.9, 8.4, 5.3)\,\Msun\ respectively, 
yielding M\textsubscript{lower} $\simeq$\,69$^{+34}_{-18}$\,/Msun. 

In rotating stars the pulsational instability region is shifted by the
additional centrifugal force. This causes the BH masses to increase,
depending on the efficiency of angular momentum transport.
\citet{marchant_2020_aa} showed that under assumptions that maximize
the impact of rotation on the gap (no angular momentum transport and
fast rigid rotation at He ignition), the shift in the peak BH mass is
$\lesssim 10\%$, similar to the findings of \citet{woosley_2021_aa}.

Assuming GW190521 formed as an isolated M\,=\,95.3\,\Msun\ BH implies an S-factor of
$\gtrsim$\,77\,keV\,b (or 88\,keV\,b with an enhanced 3-$\alpha$ rate). This is consistent with the 73$^{+11}$\,keV\,b 
inferred by \citet{farmer_2020_aa}, which used the \citet{kunz_2002_aa} $^{12}$C($\alpha$,$\gamma$)$^{16}$O reaction rate. 
The next most massive BH, GW190929 with M\,=\,80.8\,\Msun,
implies an S-factor $\gtrsim$\,98\,keV\,b (or 119\,keV\,b for an enhanced 3-$\alpha$ rate).
This is consistent with \citet{aadland_2022_aa} who suggest that observations of WO-type Wolf Rayet stars 
are best matched by models with a 25--50\% reduced $^{12}$C($\alpha$,$\gamma$)$^{16}$O rate.

\section{Conclusions}\label{s.conc}
Three physics driven transitions in the BH initial mass function are predicted by single star stellar evolution. We have have focused this work on the lower edge of the upper BH mass gap, the second transition. By evolving He cores from He-ZAMS to their final fate we have explored physically motivated mixing floors, varying spatial and temporal resolutions, and a wide range of C/O core compositions to assess the numerical convergence of our \MESA\ stellar evolution models. We find:

\begin{itemize}
\item The inclusion of $D_{\rm min}$\,=\,10$^{-2}$ cm\textsuperscript{2}/s improves the rate of convergence of models run with lower temporal resolution,
but cannot fully recover the convective behavior generated by models which use higher temporal resolutions.

\item The mass resolution needed to resolve the peak of the BH mass spectrum is limited by the amount temporal resolution available. Models with high mass and low temporal resolution smear out the BH mass spectrum peak. Models with lower mass resolution and higher temporal resolution display the largest overall peak BH mass. In \MESA\, timesteps limited by $\delta$\textsubscript{log$\rho_{c}$}$\leq 2.5\times10^{-4}$ are needed to resolve the peak BH mass to within $\leq$\,1\,\Msun, and $\delta$\textsubscript{log$\rho_{c}$}$\leq 10^{-4}$ are needed to accurately resolve the shape of the BH mass spectrum.

\item By adopting a resolution of \code{5m\_2h\_10t} or \code{5m\_2h\_10t\_D} our peak BH mass models are robust to mass and temporal resolution at the $\simeq$\,1\% level at $\sigma$[$^{12}$C($\alpha$,$\gamma$)$^{16}$O]\,=\,0 and $\simeq$\,2\% level across the mass gap. This resolution underestimates the peak BH mass by $\leq$\,2\,\Msun\ at $\sigma$\,=\,$-$3 and overestimate the peak BH mass by $\leq$\,1\,\Msun\ at $\sigma$\,=\,3. The resolved lower edge shows reasonable agreement with the results of \citet{mehta_2022_aa} at the $\leq$\,4\,\Msun\ level across the lower edge of the BH mass gap.  We establish a new lower edge of the upper mass gap as M\textsubscript{lower} $\simeq$\,60$^{+32}_{-14}$\,\Msun\ from the $\pm 3\sigma$ uncertainty in the $^{12}\text{C}(\alpha, \gamma) ^{16}\text{O}$ reaction rate probability distribution function \citep{mehta_2022_aa}.

\item  At $\sigma$[$^{12}$C($\alpha$,$\gamma$)$^{16}$O]\,=\,$-$3, the BH mass spectrum grows linearly with initial He core mass until models undergo PISNe. If the true $^{12}$C($\alpha$,$\gamma$)$^{16}$O rate is in fact this weak, we do not expect to observe any Z\,=\,10$^{-5}$ PPISN stars as the lower edge of the pair instability strip is synchronous with the lower edge of the BH mass gap.  At $\sigma$[$^{12}$C($\alpha$,$\gamma$)$^{16}$O]\,=\,0, the lower edge of the pair instability strip is M\textsubscript{He} $\sim$\,47\,\Msun. At $\sigma$[$^{12}$C($\alpha$,$\gamma$)$^{16}$O]\,=\,+3, our highest resolution models (\code{2m\_2h\_25t}) indicate the lower edge of the pair instability strip could be as low as M\textsubscript{He} $\simeq$\,32\,\Msun, about 5\,\Msun\ lower than at any other resolution. High temporal resolution is necessary to resolve the lower edge of the pair-instability strip for models with low C/O mass fractions.

\item Increased temporal resolution is important during the hydrodynamic phase due to a tight coupling between the nuclear burning and time dependant convection. Models with larger $^{12}$C($\alpha$,$\gamma$)$^{16}$O rates yield cores with low C/O mass fractions which undergo stronger nuclear burning episodes and convection and experience hundreds of pulses.

\item We explored stronger 3-$\alpha$ reaction rates by implementing a 37.8\% larger rate as a multiplicative factor on the currently adopted NACRE rate. At $\sigma$[$^{12}$C($\alpha$,$\gamma$)$^{16}$O]\,=\,$-3, 0, +3$, adopting a stronger 3-$\alpha$ rate results in an increase in the lower edge of the Upper stellar mass gap of M\textsubscript{BH} = 11.9, 8.4, 5.3\,\Msun\ respectively, yielding M\textsubscript{lower} $\simeq$\,69$^{+34}_{-18}$\,\Msun.

\end{itemize}
Future efforts to explore the resolved evolution of PPISN stars and their BH mass spectrum could 
consider coupling the temperature dependant uncertainties in the 
$^{12}$C($\alpha$,$\gamma$)$^{16}$O and 3-$\alpha$ rate probability distribution functions
\citep[e.g.,][]{fields_2018_aa}, larger nuclear reaction networks, time dependent convection models \citep{kupka_2022_aa,jermyn_2022_aa}, 
and variations in the prescriptions for wind driven mass loss, angular momentum transport, convective-core overshooting \citep{vink_2021_ras,Tanikawa_2021_aa} and binary interactions.

A goal of forthcoming low-energy nuclear experiments is to further reduce the uncertainty in the $^{12}$C($\alpha$,$\gamma$)$^{16}$O reaction rate probability distribution
\citep{deboer_2017_aa,smith_2021_aa,aliotta_2022_aa}.
Partnering with this laboratory astrophysics quest are other avenues for placing 
astrophysical constraints on the $^{12}$C($\alpha$,$\gamma$)$^{16}$O reaction rate
from the period spectrum of variable carbon-oxygen white dwarfs \citep{chidester_2022_aa},
lifetimes of He core burning stars \citep{imbriani_2001_aa,jones_2015_aa},
and the surface abundances of WO-type Wolf-Rayet stars \citep{aadland_2022_aa}.

\acknowledgements

We thank James Deboer for sharing the $^{12}$C$(\alpha,\gamma)^{16}$O probability distribution function.
This research is supported by the National Science Foundation (NSF) under 
grant PHY-1430152 for the ``Joint Institute for Nuclear Astrophysics - Center for the Evolution of the Elements'', and 
by the NSF under the Software Infrastructure for Sustained Innovation
grants ACI-1663684, ACI-1663688, and ACI-1663696 for the \MESA\ Project.
This research made extensive use of the SAO/NASA Astrophysics Data System (ADS).

\software{
\MESA\ \citep[][\url{http://mesa.sourceforge.net}]{paxton_2011_aa,paxton_2013_aa,paxton_2015_aa,paxton_2018_aa,paxton_2019_aa,jermyn_2022_aa},
\texttt{MESASDK} 20190830 \citep{mesasdk_linux,mesasdk_macos},
\texttt{matplotlib} \citep{hunter_2007_aa}, and
\texttt{NumPy} \citep{der_walt_2011_aa}.
         }

\appendix

\begin{figure*}
    \centering
    \includegraphics[width=5in]{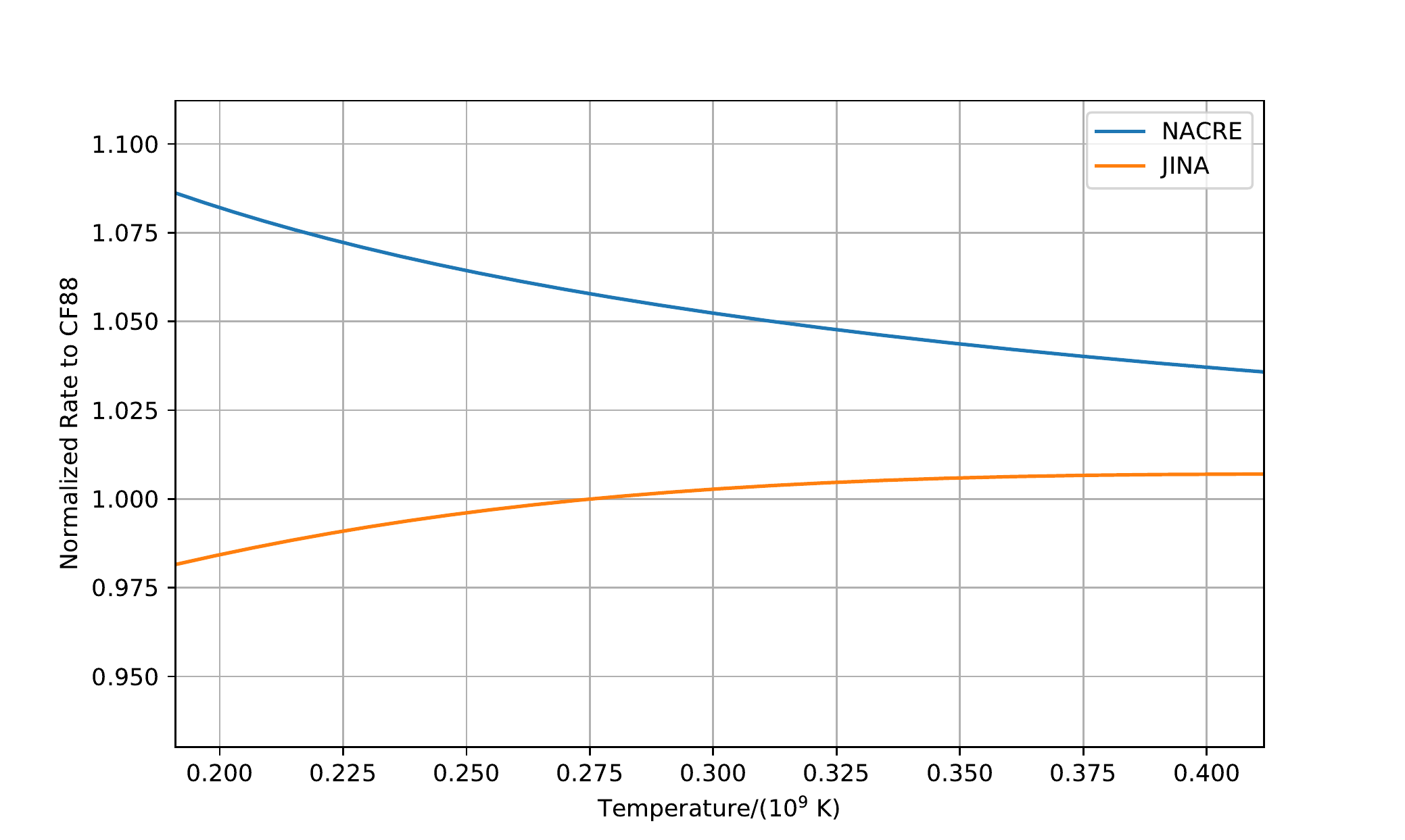}
    \caption{Normalized 3-$\alpha$ reaction rates over core He-burning temperatures 0.2 - 0.4 GK. Both NACRE and JINA rates are normalized to the 3-$\alpha$ rate from \citet{caughlan_1988_aa} (CF88).}
    \label{fig:fig8}
\end{figure*}

\MESA\ tabulates nuclear reaction rates with a grid of 10,000 temperature points evenly spaced over 0\,-\,20 GK. Figure \ref{fig:fig8} illustrates \MESA's implementation of both NACRE and JINA 3-$\alpha$ rates normalized to \citet{caughlan_1988_aa} (CF88) at He burning temperatures.


\bibliographystyle{aasjournal}

\end{document}